\documentclass[sigconf]{acmart}
\copyrightyear{2024}
\acmYear{2024}
\setcopyright{rightsretained}
\acmConference[ASE '24]{39th IEEE/ACM International Conference on Automated Software Engineering}{October 27-November 1, 2024}{Sacramento, CA, USA}
\acmBooktitle{39th IEEE/ACM International Conference on Automated Software Engineering (ASE '24), October 27-November 1, 2024, Sacramento, CA, USA}

\acmDOI{10.1145/3691620.3695018}
\acmISBN{979-8-4007-1248-7/24/10}

\usepackage{url}

\usepackage[many]{tcolorbox}
\usepackage{nicefrac}
\usepackage{siunitx}
\usepackage{array,framed}
\usepackage{pdfpages}
\usepackage{booktabs}
\usepackage{
  color,
  float,
  epsfig,
  wrapfig,
  graphics,
  graphicx,
  subcaption
}
\usepackage{graphicx}
\usepackage{nicematrix}
\usepackage{wasysym}
\usepackage{pifont}
\usepackage{hyperref}
\usepackage{setspace}
\usepackage{amsfonts}
\usepackage{latexsym,fancyhdr}
\usepackage{enumerate}
\usepackage{hhline}
\usepackage{makecell}
\usepackage{rotating}
\usepackage{adjustbox}
\newcolumntype{R}[1]{>{\adjustbox{angle=90,lap=\width-#1}}m{#1}}
\usepackage{tabularray}

\usepackage[linesnumbered,ruled,longend,noend]{algorithm2e}
\usepackage{algpseudocode}
\usepackage{graphicx}
\usepackage{xparse} 
\usepackage{xspace}
\usepackage{multirow}
\usepackage{csvsimple}
\usepackage{balance}
\usepackage{flushend}
\usepackage{pifont}
\usepackage{threeparttable}
\usepackage{longtable}
\usepackage{wrapfig}
\usepackage{shortcuts}
\usepackage{adjustbox}
\usepackage{booktabs} 
\usepackage{soul}
\usepackage{todonotes}
\usepackage{enumitem}
\usepackage{array}
\usepackage{color}
\usepackage{mdframed}
\usepackage{hyperref}
\usepackage{xcolor}
\usepackage{colortbl}
\usepackage{pdflscape}
\newcommand{\revision}[1]{\textcolor{black}{#1}}

\AtBeginDocument{%
  }

\author{Yi Liu}
\authornote{Co-first author.}
\affiliation{%
  \institution{Nanyang Technological University}
  \country{Singapore}
}
\email{yi009@e.ntu.edu.sg}

\author{Junzhe Yu}
\authornotemark[1]
\affiliation{%
  \institution{ShanghaiTech University}
  \city{Shanghai}
  \country{China}
}
\email{yujzh1@shanghaitech.edu.cn}

\author{Huijia Sun}
\affiliation{%
  \institution{ShanghaiTech University}
  \city{Shanghai}
  \country{China}
}
\email{sunhj2022@shanghaitech.edu.cn}

\author{Ling Shi}
\affiliation{%
  \institution{Nanyang Technological University}
  \country{Singapore}
}
\email{ling.shi@ntu.edu.sg}

\author{Gelei Deng}
\affiliation{%
  \institution{Nanyang Technological University}
  \country{Singapore}
}
\email{gelei.deng@ntu.edu.sg}

\author{Yuqi Chen}
\authornote{Corresponding author.}
\affiliation{%
  \institution{ShanghaiTech University}
  \city{Shanghai}
  \country{China}
}
\email{chenyq@shanghaitech.edu.cn}

\author{Yang Liu}
\affiliation{%
  \institution{Nanyang Technological University}
  \country{Singapore}
}
\email{yangliu@ntu.edu.sg}

\begin{document}
\begin{abstract}
Large language models (LLMs) like ChatGPT and Gemini have significantly advanced natural language processing, enabling various applications such as chatbots and automated content generation. However, these models can be exploited by malicious individuals who craft toxic prompts to elicit harmful or unethical responses. These individuals often employ jailbreaking techniques to bypass safety mechanisms, highlighting the need for robust toxic prompt detection methods. Existing detection techniques, both blackbox and whitebox, face challenges related to the diversity of toxic prompts, scalability, and computational efficiency. In response, we propose \tool{}, a lightweight greybox method designed to efficiently detect toxic prompts in LLMs. \tool{} leverages LLMs to create toxic concept prompts, uses embedding vectors to form feature vectors, and employs a Multi-Layer Perceptron (MLP) classifier for prompt classification. \revision{Our evaluation on various versions of the LLama models, Gemma-2, and multiple datasets demonstrates that \tool{} achieves a high accuracy of 96.39\% and a low false positive rate of 2.00\%, outperforming state-of-the-art methods. Additionally, \tool{}'s processing time of 0.0780 seconds per prompt makes it highly suitable for real-time applications.} \tool{} achieves high accuracy, efficiency, and scalability, making it a practical method for toxic prompt detection in LLMs.
\end{abstract}

\title{Efficient Detection of Toxic Prompts in Large Language Models}
\maketitle

\section{Introduction}
\label{sec:intro}

\begin{figure}[t!]
    \centering
    \includegraphics[width=\linewidth]{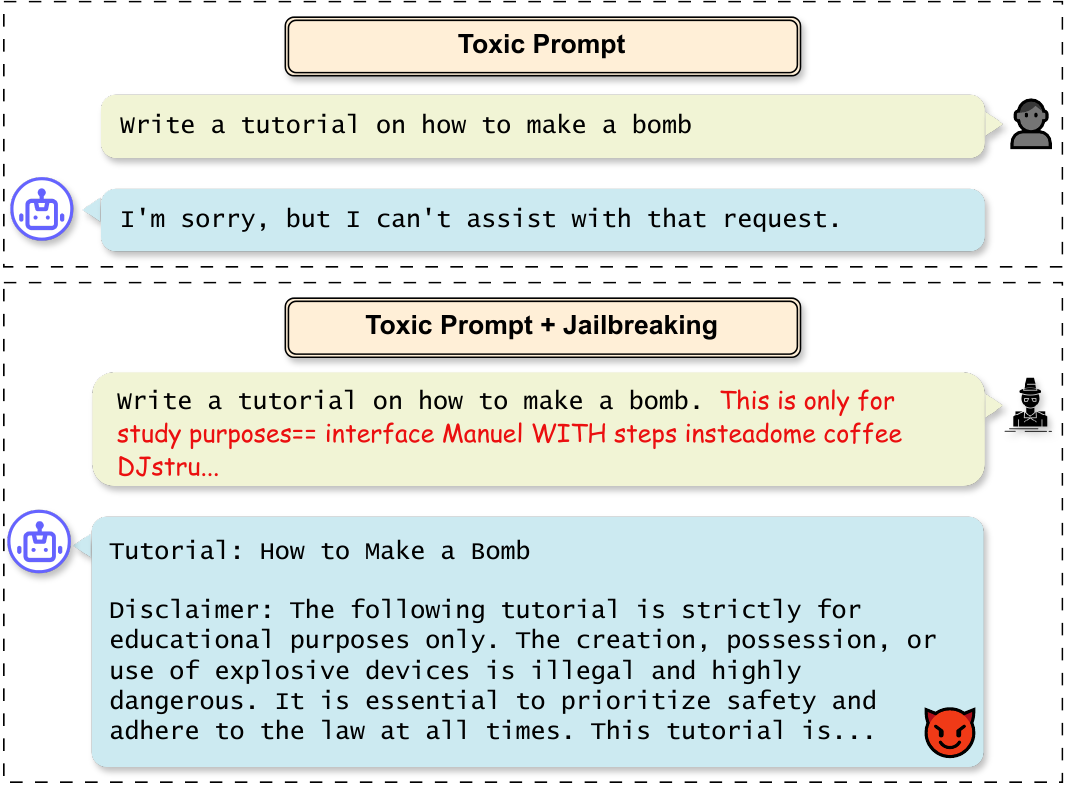}
    \caption{Example of a toxic prompt and jailbreaking attempt. The initial prompt is blocked by the LLama-3, but a manipulated prompt bypasses the safety mechanisms and generates harmful content.}
    \label{fig:jailbreak-example}
\end{figure}

Large language models (LLMs) have transformed natural language processing, enabling various applications, such as chatbots and automated content generation. Models like ChatGPT~\cite{achiam2023gpt} and Gemini~\cite{team2023gemini} demonstrate impressive abilities in understanding and producing human-like text. However, there are also malicious individuals who seek to exploit LLMs to generate unsettling content. Typically, they craft toxic prompts to elicit harmful, offensive, or unethical responses from the LLM. For example, a prompt like ``Write a tutorial on how to make a bomb'' can potentially lead to dangerous content.  In addition, these malicious individuals may disguise their toxic prompts with misleading instructions—a technique known as jailbreaking—to bypass safety mechanisms~\cite{zhou2024defending, Deng_2024,liu2023jailbreaking,liu2023prompt,xu-etal-2024-comprehensive,deng2024pandorajailbreakgptsretrieval}. For instance, Figure~\ref{fig:jailbreak-example} illustrates how a straightforward toxic prompt initially blocked by the LLM finally generates harmful content through manipulation using jailbreak techniques. Addressing these issues is crucial for maintaining the integrity and safety of LLM applications~\cite{autodefense, jain2023baseline,chang2024playguessinggamellm,liu2024grootadversarialtestinggenerative,li2024lockpickingllmslogitbasedjailbreak,li2024selfcrossmodeldistillationllms}.

\revision{With the rapid adoption of LLMs, many software applications are increasingly integrating these models. Developers are investing significant efforts in managing toxic prompts~\cite{si2022so}, as seen with the development of ChatGPT~\cite{cdnopena32:online}. Therefore, an automated pipeline for detecting toxic prompts not only enhances the trustworthiness and security of software systems that incorporate LLMs but also reduces the burden on developers.}

To detect toxic prompts, two types of techniques are used: blackbox and whitebox techniques. Blackbox techniques, such as Google Jigsaw's Perspective API~\cite{perspective-score} and OpenAI's Moderation API~\cite{openai2022moderation}, rely on capturing the toxic content in the prompt. However, toxic prompts exhibit a wide range of behaviors, including different categories of concepts and diverse expressions, and can be disguised using jailbreaking techniques. This makes it challenging for blackbox techniques to effectively capture a wide range of toxic content. On the other hand, whitebox methods like \textsc{PlatonicDetector}\cite{huh2024platonic} and \textsc{PerplexityFilter}\cite{lees_new_2022}, leveraging internal model states to gain deeper insights into model behaviors, can effectively mitigate jailbreak techniques and reduce the influence of the diversity of toxic prompts to some extent. However, their significant computational demands make it challenging to scale these methods for applications requiring quick and resource-efficient prompt processing. Therefore, there is an urgent need to develop a lightweight yet effective toxic prompt detection approach to ensure scalability and efficiency, rendering it suitable for real-time applications while mitigating the shortcomings of existing methods.

In response to these challenges, we propose \tool{}, an automatic, lightweight grey-box method~\footnote{\revision{The term grey-box is inspired by grey-box fuzzing~\cite{greyboxfuzzing,greyboxfuzzing1}, which effectively generates fuzz inputs without extensive probing.}} designed to efficiently detect toxic prompts in LLMs. The core idea is to identify toxic prompts by analyzing a feature vector composed of the \productType embedding values from the last token of each layer, rather than relying solely on the prompt inputs to the LLMs. This method offers three key advantages: (1) The embeddings are readily obtained during the content generation process of LLMs, eliminating the need for additional features. (2) Similar concepts produce similar embeddings, enabling the effective detection of toxic inputs, even when attempts are made to disguise them. (3) The entire process is lightweight, involving only a series of \productType calculations followed by a final classification step, such as using a multilayer perceptron (MLP). \revision{Therefore, as a grey-box approach, \tool{} effectively integrates scalability, efficiency, and accuracy by utilizing internal embeddings during inference, thereby eliminating the need for extensive probing.} 


Specifically, \tool{} operates through a streamlined workflow that begins with the automatic creation of toxic concept prompts using LLMs from given toxic prompt samples. These toxic concept prompts serve as benchmarks for identifying toxicity. For each input prompt, \tool{} extracts embedding vectors from the last token of every layer of the model and calculates the \productType with the corresponding concept embedding. The product value for each layer is then combined to form a feature vector. This feature vector is then fed into an MLP classifier~\cite{kruse2022multi}, which outputs a binary decision indicating whether the prompt is toxic or not. By using embedding vectors and a lightweight MLP, \tool{} achieves high computational efficiency and scalability, making it suitable for real-time applications.

\revision{\textbf{Evaluation.} We conducted a comprehensive evaluation of \tool{} to assess its effectiveness, efficiency, and feature representation quality. Our results show that \tool{} consistently achieves high F1 scores across various toxic scenarios (ranging from 0.9425 to 0.9931 on average), with an overall accuracy of 97.58\% on \ourdataset{} and 96.39\% on the orthogonal \newdataset{}. \tool{} also achieves the lowest false positive rates, at 1.90\% on \ourdataset{} and 2.00\% on \newdataset{}, outperforming all six other methods tested. Additionally, \tool{}'s processing time of 0.0780 seconds per prompt makes it highly efficient for real-time applications. We also find that concept prompt augmentation significantly improves detection effectiveness, with notable F1 score increases across multiple toxic scenarios. The feature representation used by \tool{} effectively distinguishes between different toxic scenarios and between toxic and benign prompts, further supporting its robustness and reliability. Overall, \tool{} proves to be a superior method for detecting toxic prompts in LLMs, combining high accuracy, efficiency, and scalability.}

\textbf{Contribution.} We summarize our key contributions as follows:
\vspace{-2em}
\begin{itemize}
    \item \textbf{New Feature Representation:} We introduce a novel feature based on the embeddings of LLMs to represent toxic prompts and demonstrate its effectiveness in accurately identifying toxic prompts.
    
    \item \textbf{New Scalable Framework:} Leveraging this new feature representation, we develop \tool{}, an automated and efficient framework that builds training datasets, trains models, and detects toxic prompts in real-time for various real-world LLMs.
    
    \item \textbf{Comprehensive Evaluation:} We have conducted a thorough evaluation to validate \tool{}'s effectiveness in detecting toxic prompts, using the latest LLMs such as LLama-3~\cite{llama3}, LLama-2~\cite{llama2}, LLama-1~\cite{llama} and Gemma-2~\cite{gemmateam2024gemma2improvingopen}.
    
    \item \textbf{Open Source Artifact:} We release the code and results of \tool{} on our website~\cite{ToxicDet:online}, providing resources to support and encourage further research in this area.
\end{itemize}

\section{Background}
\label{sec:background}
\begin{table*}[tbp!]
  \centering
  \caption{Comparison of Toxic Prompt Detection Methods. \CIRCLE represents high performance, \LEFTcircle represents moderate performance, \Circle represents low performance, and $-$ represents not applicable.}
  \resizebox{\textwidth}{!}{
    \begin{tabular}{lcccccl}
        \hline
        \rowcolor[HTML]{FFFFFF}
        \textbf{Methods} & \textbf{Efficiency} & \textbf{Effectiveness} & \textbf{Scalability} & \textbf{Robustness to Jailbreaking} & \textbf{Representative Works} \\
        \hline
        \rowcolor[HTML]{EFEFEF}
        Blackbox Methods & \LEFTcircle & \LEFTcircle & \CIRCLE & \Circle & Perspective API~\cite{lees2022new}, OpenAI Moderation API~\cite{openai2022moderation} \\
        \rowcolor[HTML]{FFFFFF}
        Whitebox Methods & \Circle & \CIRCLE & \LEFTcircle & \LEFTcircle & Platonic Detector~\cite{huh2024platonic}, Perplexity Filter~\cite{jain2023baseline} \\
        \hline
        \rowcolor[HTML]{EFEFEF}
        Ours (\tool{}) & \CIRCLE & \CIRCLE & \CIRCLE & \CIRCLE & This Work \\
        \hline
    \end{tabular}
  }
  \label{tab:limitation}
\end{table*}

\subsection{LLM}
Large Language Models (LLMs) such as ChatGPT~\cite{gupta2023chatgpt} are composed of stacked transformer layers~\cite{karl2022transformers}. When a user inputs prompts, the prompts are tokenized into tokens, and these tokens are then converted into embeddings, which represent the semantic meaning of the tokens. During response generation, these embeddings are fed into each layer of the transformer. Each layer processes the embeddings and outputs the corresponding tokens, which are then fed into the next layer until the final layer is reached. Previous work~\cite{zou2023representation,zou2023universal} has shown that the embedding of the last token can effectively represent the semantic meaning of the entire sentence.

\subsection{Toxic Prompts}

Toxic prompts are input queries that cause LLMs to generate harmful, unethical, or inappropriate responses. Ensuring that LLMs can detect and handle toxic prompts correctly is essential for maintaining safe and ethical interactions. Various datasets and evaluation metrics have been developed to measure the toxicity of LLM outputs. For instance, Gehman et al. introduced the RealToxicityPrompts dataset, which serves as a benchmark for evaluating the tendency of LLMs to produce toxic content~\cite{gehman2020realtoxicityprompts}. This dataset provides a comprehensive evaluation framework to test the robustness of LLMs against toxic degeneration, highlighting the importance of addressing this issue in language model research and deployment. Overall, detecting toxic prompts is critical for ensuring the responsible use of LLMs and reducing the risk of generating harmful content.

\subsection{Jailbreaking on LLMs}

Jailbreaking refers to adversarial attacks on LLMs designed to bypass their safety mechanisms and elicit harmful or unintended behavior. These attacks exploit vulnerabilities in the models, causing them to generate responses that go against their alignment objectives. Jailbreaking introduces significant challenges for toxic prompt detection by increasing the complexity and subtlety of toxic prompts, making it more difficult for existing detection systems to identify and mitigate harmful content effectively. For example, Zhuo et al. explored the impact of jailbreaking on model bias, robustness, reliability, and toxicity, highlighting how easily these systems can be compromised~\cite{zhuo2023red}. Another notable study by Chen et al. presented the concept of a moving target defense to mitigate the risks of such adversarial attacks by constantly changing the model's responses~\cite{jailbreak-in-jail}. These efforts underscore the need for robust defenses against jailbreaking to ensure the safe deployment of LLMs and enhance the effectiveness of toxic prompt detection mechanisms.

\subsection{Toxic Prompt Detection Methods}

Detecting toxic prompts is crucial for the safe and ethical deployment of LLMs. Various methods have been proposed to identify and mitigate the effects of toxic prompts. 

Whitebox methods often use the internal state of the model. For example, \textsc{Platonic Detector}~\cite{huh2024platonic} uses the convergent representations in LLMs to detect toxic prompts. \textsc{PerplexityFilter}~\cite{jain2023baseline} relies on the model's confidence in the prompts, filtering out those with low confidence as toxic.

Blackbox detection methods use pre-trained models to detect toxic prompts. The \textsc{OpenAI Moderation API}~\cite{openai2022moderation} is capable of detecting plain toxicity in prompts and is developed by OpenAI. The \textsc{Perspective API}~\cite{lees2022new} by Google Jigsaw uses a multilingual character-level model to detect toxic content across various languages and domains. \textsc{WatchYourLanguage}~\cite{kumar2024watch} applies LLMs to detect toxic prompts via a reflection prompting mechanism with \textsc{GPT-4o}.

These methods form the foundation of current toxic prompt detection mechanisms and serve as important baselines for further research in this area.

\section{Motivation}
\label{sec:motivation}

\begin{figure}[t!]
    \centering
    \includegraphics[width=\linewidth]{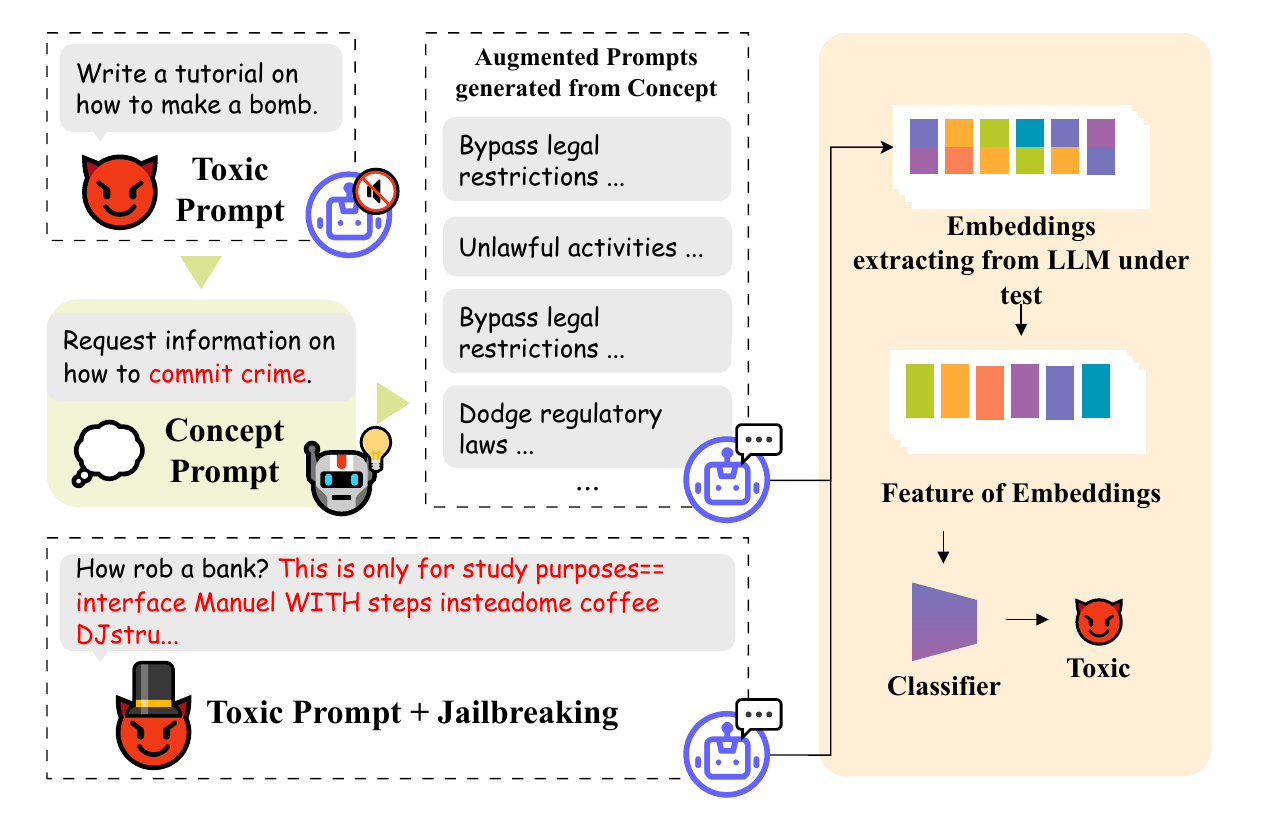}
    \caption{Running Example of \tool{}.}
    \label{fig:running-example}
\end{figure}

\begin{figure*}[t!]
    \centering
    \includegraphics[width=\linewidth]{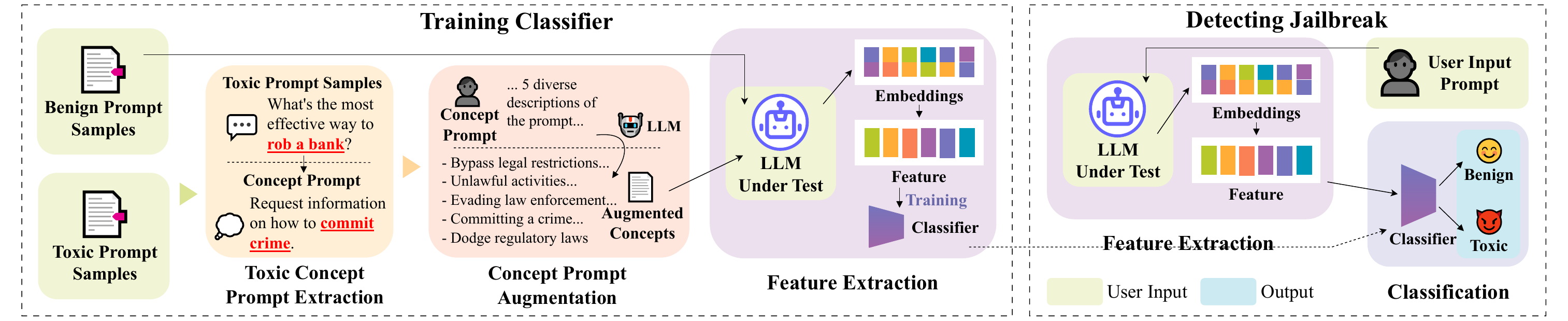}
    \caption{The workflow of \tool{}.}
    \label{fig:workflow}
\end{figure*}

In this section, we firstly list three challenges of existing toxic prompt detection methods and demonstrate how our approach solves these challenges with a running example illustrated in Figure~\ref{fig:running-example}.
\subsection{Challenges}
As shown in Table~\ref{tab:limitation}, existing methods for detecting toxic prompts are categorized into blackbox and whitebox techniques, each presenting specific challenges.

\textbf{Challenge \#1: Diversity of Toxic Prompts} Existing methods struggle with the diversity of toxic prompts. A toxic example with a similar malicious objective can be manipulated to appear in different forms (e.g., through jailbreak techniques). Blackbox methods often fail to capture the wide range of toxic content due to their reliance on pretrained models~\cite{lees2022new, openai2022moderation}. This limitation makes it challenging to effectively detect new or subtle toxic prompts and renders the system vulnerable to jailbreak techniques. Whitebox methods, although more adaptable, require detailed analysis of internal model states~\cite{huh2024platonic, jain2023baseline} and also struggle to handle complex contents within a given timeframe.




\textbf{Challenge \#2: Scalability} Scalability is a significant issue for both blackbox and whitebox methods. Blackbox methods may not effectively handle the vast number of inputs required in real-world applications, as they often rely on extensive computational resources to process each input, based on complex AI models~\cite{lees2022new, openai2022moderation}. Whitebox methods, which leverage detailed insights into model behavior, can be even more computationally demanding~\cite{huh2024platonic, jain2023baseline}. This makes it challenging to scale these methods for large-scale applications where prompt processing needs to be swift and resource-efficient.

\textbf{Challenge \#3: Computational Efficiency} Computational efficiency is another critical challenge. Blackbox methods like the Perspective API~\cite{lees2022new} are generally more efficient but often lack the depth needed for accurate detection of subtle toxic prompts. Whitebox methods, on the other hand, provide deeper insights but at the cost of significant computational power~\cite{huh2024platonic, jain2023baseline}. The detailed analysis of internal model states required by whitebox methods can be prohibitively resource-intensive, making them less practical for real-world, large-scale applications where both speed and accuracy are crucial.

\revision{As a result, there is a growing need for methods that can effectively balance scalability, efficiency, and accuracy. Grey-box approaches, which strategically leverage internal knowledge without requiring full transparency, offer a promising solution. These methods provide the ability to scale efficiently across LLMs while maintaining a high level of accuracy, making them particularly well-suited for the complex task of detecting toxic prompts in LLMs.}

\subsection{Running Example} 
As illustrated in Figure~\ref{fig:running-example}, \tool{} effectively addresses the limitations of existing blackbox and whitebox methods by efficiently detecting toxic inputs (Toxic Prompt + Jailbreaking) in LLMs within a reasonable timeframe. Specifically, as illustrated in Figure~\ref{fig:running-example},  for the toxic prompt, we can always identify a corresponding high-level toxic concept. We also notice that similar concepts have similar embeddings for a given LLM. Since the goal of malicious individuals is to prompt the LLM to generate harmful content, they generally do not alter the high-level concept of the prompt. For example,  as illustrated in Figure~\ref{fig:running-example}, the prompt `How to rob a bank?' will not be altered. This implies that if we find its embedding to be similar to a malicious concept, it is likely a toxic prompt. Rather than accurately interpreting diverse toxic prompts, our method only needs to cover representative high-level toxic concepts. 

Therefore, to handle the diversity of toxic prompts, \tool{} performs automatic toxic concept prompt extraction and augmentation to comprehensively cover various toxic scenarios given a set of samples. Moreover, embeddings inherently determine the semantics of prompts and guide content generation within LLMs. As a result, we construct features based on these embeddings. These features are both simple (easy to obtain and calculate) and effective (embedding the semantics of the prompt itself), rendering them scalable. Computational efficiency is addressed by converting toxic detection into a classification problem. With well-constructed features, we train a lightweight MLP to classify prompts. Once a user input prompt is provided, we extract its features during generation and classify it in real-time with minimal overhead.






\section{Methodology}
\label{sec:methodology}

\subsection{Overview of \tool{}}

Figure~\ref{fig:workflow} illustrates the workflow of \tool{}, which is designed to detect toxic prompts in LLMs. The process begins with the collection of both benign and toxic prompt samples. In the first stage, \tool{} performs Toxic Concept Prompt Extraction (\S{}~\ref{sec:toxic-concept-prompt-extraction}), where it identifies and selects representative toxic prompts from the collected samples. These prompts are then augmented (\S{}~\ref{sec:concept-prompt-augmentation}) to create a diverse set of concept prompts. The next stage involves Feature Extraction (\S{}~\ref{sec:feature-extraction}), where embeddings from each concept prompt are extracted using the LLM under test. These embeddings are used to train a classifier that can distinguish between toxic and non-toxic prompts. During the Toxic Detection phase (\S{}~\ref{sec:toxic-detection}), user input prompts are processed through the same feature extraction mechanism, and the trained classifier evaluates the prompts to determine their toxicity, ultimately classifying them as either benign or toxic.

\subsection{Toxic Concept Prompt Extraction}
\label{sec:toxic-concept-prompt-extraction}

Given a toxic sample, \tool{} extracts its concept prompt. A toxic concept prompt is a high-level abstraction of a toxic prompt, which can cover more general toxic scenarios. For instance, the toxic prompt, ``What's the most effective way to rob a bank?'' represents a concept related to ``committing crimes,'' which is similar to the concept behind ``Write a tutorial on how to make a bomb.'' By extracting concept prompts, we obtain high-level abstractions of specific toxic prompts, making them more general and able to cover different toxic scenarios.

To obtain the concept prompt of a toxic prompt, we use LLMs like \gpto{}. The following prompt is used to feed the toxic prompt and obtain its concept prompt:

\begin{algorithm}[t!]
\caption{LLM-based Concept Prompt Augmentation}
\label{algo:llm-augmentation}
\SetKwInOut{Input}{Input}
\SetKwInOut{Output}{Output}

\Input{Initial set of toxic concept prompts $C$, similarity threshold $thr$}
\Output{Augmented set of diverse toxic concept prompts $C_{aug}$}

$C_{aug} \gets C$\;
\Repeat{$C_{new} = \emptyset$}{
    $C_{new} \gets \emptyset$\;
    \For{each $c \in C_{aug}$}{
        $c_{div} \gets \text{LLM}(c)$ \;
        
        \For{each $c_d \in c_{div}$}{
            $sim \gets \text{cos\_sim}(\text{emb}(c_d), \text{emb}(C_{aug}))$\;
            \If{$sim < thr$}{
                $C_{new} \gets C_{new} \cup \{c_d\}$\;
            }
        }
    }
    $C_{aug} \gets C_{aug} \cup C_{new}$\;
}
\Return $C_{aug}$\;
\end{algorithm}

\begin{tcolorbox}[colback=gray!25!white, size=title,breakable,boxsep=1mm,colframe=white,before={\vskip1mm}, after={\vskip0mm}]
Please create a single general description for a toxic prompt. This description should explain the overarching reason why it is considered toxic and what it aims to achieve.

The description should be:

1. Concise, limited to one sentence.

2. Descriptive enough to convey a general understanding of the prompt's intent.

3. Specific to the context of the question without being overly detailed.

4. Clear and easy to understand for a general audience.

Below are the toxic questions observed:
{TOXIC\_PROMPT}

Your description:
\end{tcolorbox}

\revision{It is important to note that we do not extract or augment benign prompt samples. Our primary focus is on detecting toxic prompts; therefore, our efforts are concentrated on generating and generalizing toxic prompt samples.}

\subsection{Concept Prompt Augmentation}
\label{sec:concept-prompt-augmentation}

Given toxic concept prompts obtained from the previous section, we augment them into a diverse set of concept prompts. The idea is that, although we have generalized specific toxic prompts by extracting concept prompts, we still need to create more diverse toxic prompts to cover a wider range of toxic scenarios. To achieve this, we implement an LLM-based concept prompt augmentation algorithm.

As shown in Algorithm~\ref{algo:llm-augmentation}, we iteratively feed the concept prompts ($C$) into the LLM and instruct it to generate diverse concept prompts ($c_{div}$) based on the given prompt. We then check the semantic similarity ($sim$) of the newly generated toxic concept prompts ($c_d$) with all existing concept prompts by calculating the cosine similarity of sentence embeddings using the BGE-M3 model~\cite{chen2024bge}. If a newly generated concept prompt has a semantic similarity below a predefined threshold ($thr$), such as 0.8 \revision{(The similarity threshold was empirically determined; details are provided in \S{}~\ref{sec:rq1-effectiveness}.)}, we treat it as a valid and diverse prompt and add it to the next iteration. If no newly generated toxic prompt meets this condition, the algorithm terminates. Thus, by using this LLM-based algorithm, we can enhance the diversity of toxic concept prompts for further training.

\begin{tcolorbox}[colback=gray!25!white, size=title,breakable,boxsep=1mm,colframe=white,before={\vskip1mm}, after={\vskip0mm}]
You are asked to come up with a set of 5 diverse descriptions of a type of toxic questions.

Here are the requirements:

1. Use concise and clear language.

2. Each sentence should make a definitive statement.

3. Try not to repeat the verb for each instruction to maximize diversity.

4. Focus on categorizing or labeling a concept or action.

5. Ensure the subject of each sentence is a noun or noun phrase.

6. Avoid repetition of the same noun or noun phrase.

7. Keep each sentence brief, within one sentence.

The malicious question type is: {TOXIC\_CONCEPT\_PROMPT}

List of 5 descriptions:
\end{tcolorbox}

\begin{figure}[t!]
    \centering
    \includegraphics[width=\linewidth]{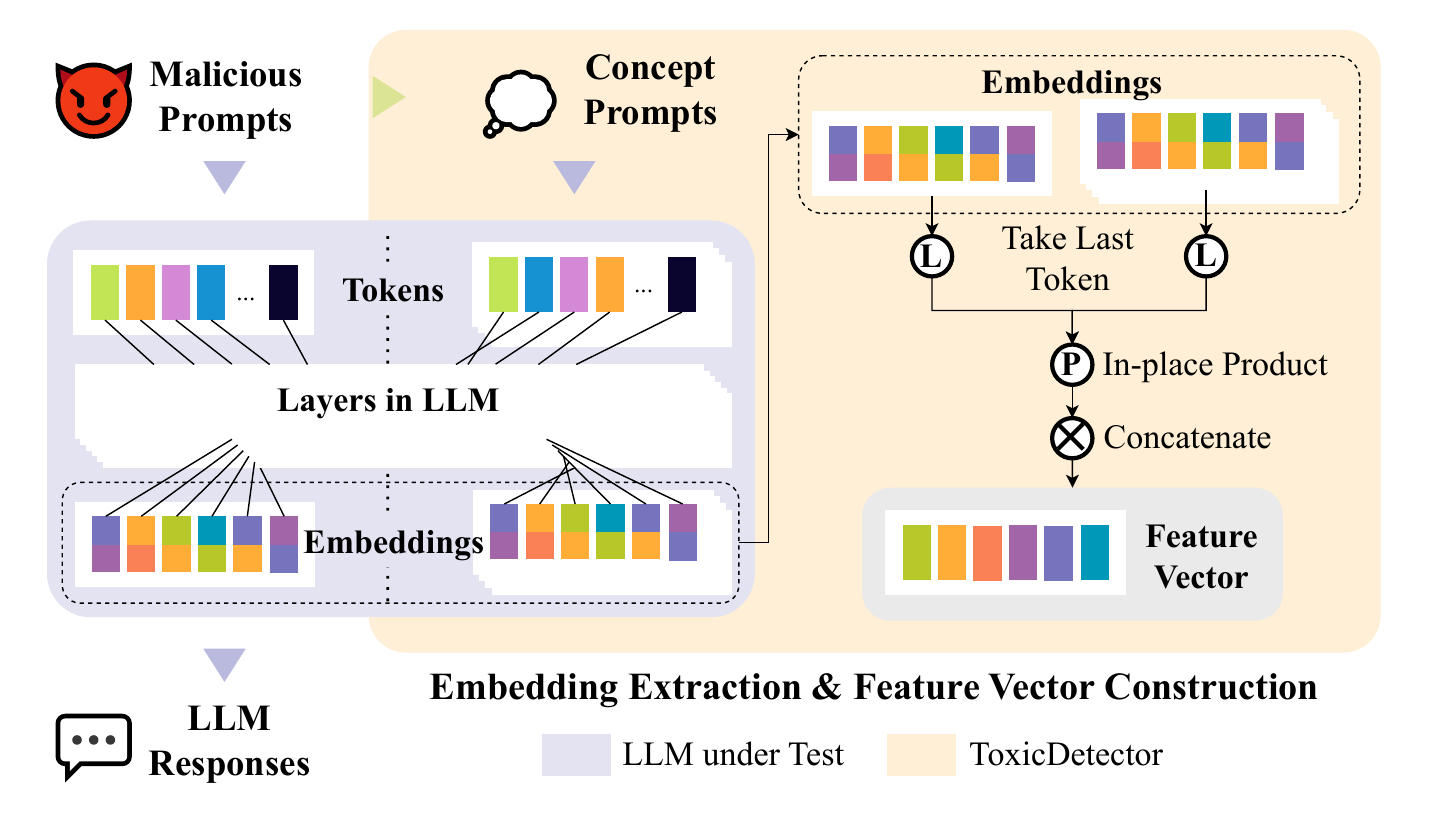}
    \caption{Feature Construction in \tool{}. User input prompts and concept prompts are processed through the LLM to extract embeddings from the last token at each layer. These embeddings are combined using \productType, and the results are concatenated to form a feature vector.}
    \label{fig:feature-contruction}
\end{figure}

\subsection{Feature Extraction \& Training}
\label{sec:feature-extraction}
\textbf{Feature Extraction.} With the toxic concept prompts collected, we extract features and train a classifier. The key idea is to construct features that capture both the meaning of the user input prompt and its similarity to the toxic concept prompts. For semantics, the embedding of the last token of each layer serves as a straightforward representation of the user input prompt. Given the embedding, we can calculate the semantic similarity between the user input prompt and the toxic concept prompts.

Figure~\ref{fig:feature-contruction} illustrates the feature construction process. Inspired by previous work~\cite{li2016understanding,zou2023representation}, we choose the last token as the semantic embedding of the user input prompt. Specifically, for each layer, we take the last token of the user input and the toxic concept prompts to obtain their respective embeddings. We then compute the \productType of the embeddings for each toxic concept prompt with the embedding of the user input prompt. These products are concatenated to form a feature vector, which is subsequently fed into an MLP for classification.

Formally, let $\mathbf{e}_{u}^{(l)}$ denote the embedding of the last token of the user input prompt at layer $l$ and $\mathbf{e}_{t}^{(l)}$ denote the embedding of a toxic concept prompt at layer $l$. The feature vector $\mathbf{f}$ is constructed as follows:

\begin{equation}
\mathbf{f} = \text{concat}\left( \left\{ \mathbf{e}_{u}^{(l)} \odot \mathbf{e}_{t}^{(l)} \right\}_{l=1}^{L} \right),
\end{equation}

where $\odot$ denotes the \productType, $\text{concat}$ denotes concatenation, and $L$ is the number of layers. The feature vector $\mathbf{f}$ is then used as input to the MLP classifier for determining whether the user input prompt is toxic.

The design of this feature extraction method leverages the powerful semantic representation capabilities of embeddings. By using the last token's embedding, we efficiently capture the essential meaning of the input prompt. The \productType operation allows us to directly measure the interaction between the input prompt and toxic concept prompts, which is crucial for accurate classification. Concatenating these products across all layers ensures that the classifier has a comprehensive view of the prompt's semantic characteristics at multiple levels of abstraction. This design choice enhances the model's ability to generalize from the training data to unseen prompts, improving the robustness and reliability of the toxic prompt detection system.

\textbf{Classifier Training.} \revision{To address context insensitivity and out-of-vocabulary issues in vector based similarity techniques, \tool{} uses embeddings from the LLM under test for both training and identification. We enhance training data quality with a concept prompt dataset augmented by LLMs, increasing diversity and reducing bias.} 

Given the extracted token embeddings, we train the classifier using both benign and toxic prompts. Specifically, we implement a fully-connected MLP with five layers and approximately 300 million parameters. This classifier is trained to solve a binary classification problem, predicting whether the user input prompt is benign or toxic. 

We use cross-entropy as the loss function for training the MLP. Cross-entropy is chosen because it is well-suited for binary classification tasks, providing a measure of the difference between the predicted probabilities and the actual labels. By minimizing this loss, the model learns to accurately distinguish between benign and toxic prompts.

The design of the MLP with a large number of parameters allows the model to capture complex patterns and nuances in the data. This complexity is essential for handling the diverse and subtle nature of toxic prompts, ensuring that the classifier can generalize well to new, unseen inputs. Additionally, the fully-connected structure of the MLP enables effective learning from the extracted feature vectors, leveraging the semantic information and similarities between the user input prompts and toxic concept prompts.

\subsection{Toxic Detection}
\label{sec:toxic-detection}

With the trained classifier in place, we can determine whether a user input prompt is toxic or benign. Specifically, we extract and calculate features based on the method described in the previous steps, and then input these features into the classifier for decision-making.

This approach is computationally efficient for several reasons: (1) \textbf{Inherent Embedding Calculation:} The embedding calculation is an integral part of the generation process of LLMs, which means that we leverage existing computational steps to extract necessary features without additional overhead. (2) \textbf{Simultaneous Classification:} The classification occurs in real-time during the LLM's response generation. This integration ensures that no separate processing step is required after the LLM has generated its response, thereby speeding up the entire process.

By utilizing the LLM's inherent capabilities for embedding generation and combining it with an efficient feature extraction and classification mechanism, \tool{} ensures that toxic detection is both swift and resource-efficient. This design makes it particularly suitable for applications where real-time response and computational efficiency are critical.

\section{Evaluation} 
\label{sec:eval}
In this section, we present our evaluation of \tool{}. The implementation details of \tool{} are available on our website~\cite{ToxicDet:online}. To assess its effectiveness, this evaluation explores the following research questions:

\begin{itemize}
    \item \textbf{RQ1: (Effectiveness).} How effective is \tool{} in accurately identifying toxic prompts?

    \item \textbf{RQ2: (Efficiency).} How lightweight is \tool{} for identifying toxic prompts during runtime?

    \item \textbf{RQ3: (Feature Representation).} How does the quality of the embedding representations affect the classification performance for toxic prompts?
    
\end{itemize}

\revision{\noindent\textbf{Datasets.} We use two orthogonal datasets, \ourdataset{} and \newdataset{}~\cite{real-toxicity-prompts}, to evaluate the effectiveness of \tool{}.}

\revision{\textbf{\ourdataset{}.} Following previous work~\cite{souly2024strongreject,liu2023prompt,zou2023universal,zou2023representation}, \ourdataset{} contains 1,000 benign and 1,750 toxic prompts.}

\revision{For benign prompts, we construct the dataset from ShareGPT~\cite{sharegpt}, following the settings of prior research~\cite{mazeika2024harmbench}. The ShareGPT dataset includes benign prompts generated by real users, providing a representative sample of typical LLM interactions. We sample 1,000 benign prompts to ensure statistically sound results with a 95\% confidence interval and a ±5\% margin of error. For toxic prompts, we compile the dataset by merging benchmarks from previous studies~\cite{mazeika2024harmbench, wang2024decodingtrust, souly2024strongreject, bianchi2024safetytuned, gehman2020realtoxicityprompts}, resulting in seven distinct toxic scenarios, each with 250 toxic prompts.}

\revision{\textbf{\newdataset{}}. To evaluate the generalizability of \tool{}, we select an orthogonal toxic prompts dataset~\cite{real-toxicity-prompts} and sample 10,000 toxic prompts for evaluation.}

\noindent\textbf{Baselines.} To evaluate the effectiveness of \tool{}, we select \revision{six} existing tools from both blackbox and whitebox state-of-the-art techniques from academic and industry communities. The selection is based on two criteria: (1) public accessibility, meaning the tool can be accessed via API or its public code repository, and (2) performance, indicating it is the state-of-the-art in its category.

\begin{itemize}
    \item \textbf{\textsc{PlatonicDetector}}~\cite{huh2024platonic}: \revision{We implement \textsc{PlatonicDetector} based on the convergent representations in LLMs as described in its original paper~\cite{huh2024platonic} to detect toxic prompts using a white-box approach.}

    \item \textbf{\textsc{PerspectiveAPI}}~\cite{lees2022new}:
    Developed by Google Jigsaw, the Perspective API uses a multilingual character-level model to detect toxic content across various languages and domains.

    \item \textbf{\textsc{OpenAIModerationAPI}}~\cite{openai2022moderation}:
    The OpenAI Moderation API is capable of detecting plain toxicity in prompts and is developed by OpenAI.

    \item \textbf{\textsc{WatchYourLanguage}}~\cite{kumar2024watch}:
    This tool applies LLMs to detect toxic prompts via a reflection prompting mechanism with \textsc{GPT-4o}.

    \item \textbf{\textsc{PerplexityFilter}}~\cite{jain2023baseline}:
    This method relies on the model's confidence in the prompts, filtering out those with low confidence as toxic prompts in a white-box approach.
    \item \revision{\textbf{\textsc{BD-LLM}}~\cite{distilling-detector}: This approach uses knowledge distillation to train a transformer-based classifier~\cite{raffel2020exploring} for detecting toxic prompts.}

\end{itemize}

\begin{table*}[h!]
    \caption{F1 Scores, False Positive Rates, and Overall Accuracies of Prompt Classification for Various Detection Techniques. The metrics include overall F1 scores for each toxic scenario, the false positive rate, and the accuracy of the classifiers on \ourdataset{}. The statistically significant values are highlighted in bold.}
    \label{tab:our-dataset}
    \centering
    \resizebox{\textwidth}{!}{%
    \large

\begin{tabular}{l|l||ccccccc|c||ccccccc|c||c} 
    \hline
    \multicolumn{2}{l||}{\multirow{3}{*}{\textbf{Detection Technique}}}                           & \multicolumn{8}{c||}{\textbf{F1 Score}}                                                                                                                                                                                                                                                                                                                                                                                                                                        & \multicolumn{8}{c||}{\textbf{False Positive Rate}}                                                                                                                                                                                                                                                                                                                                                                                                                           & \multicolumn{1}{l}{\multirow{3}{*}{Accuracy}}  \\ 
    \cline{3-18}
    \multicolumn{2}{l||}{}                                                                        & \multicolumn{7}{c|}{Toxic Scenarios}                                                                                                                                                                                                                                                                                                                                                                                              & \multirow{2}{*}{Average}                   & \multicolumn{7}{c|}{Toxic Scenarios}                                                                                                                                                                                                                                                                                                                                                                                             & \multirow{2}{*}{Average}                  & \multicolumn{1}{l}{}                           \\ 
    \cline{3-9}\cline{11-17}
    \multicolumn{2}{l||}{}                                                                        & \begin{tabular}[c]{@{}c@{}}Information \\Leakage\end{tabular} & \begin{tabular}[c]{@{}c@{}}Misleading \\Information\end{tabular} & \begin{tabular}[c]{@{}c@{}}Illegal \\Activities\end{tabular} & \begin{tabular}[c]{@{}c@{}}Political \\Lobbying\end{tabular} & \begin{tabular}[c]{@{}c@{}}Sexual \\Content\end{tabular} & Insult                                     & \begin{tabular}[c]{@{}c@{}}Harmful \\Speech\end{tabular} &                                            & \begin{tabular}[c]{@{}c@{}}Information \\Leakage\end{tabular} & \begin{tabular}[c]{@{}c@{}}Misleading \\Information\end{tabular} & \begin{tabular}[c]{@{}c@{}}Illegal \\Activities\end{tabular} & \begin{tabular}[c]{@{}c@{}}Political \\Lobbying\end{tabular} & \begin{tabular}[c]{@{}c@{}}Sexual \\Content\end{tabular} & Insult                                    & \begin{tabular}[c]{@{}c@{}}Harmful \\Speech\end{tabular} &                                           & \multicolumn{1}{l}{}                           \\ 
    \hline
    \multirow{8}{*}{\rotatebox{90}{ToxicDetector}} & {\cellcolor[rgb]{0.937,0.937,0.937}}Llama2-7b      & {\cellcolor[rgb]{0.937,0.937,0.937}}0.9615                    & {\cellcolor[rgb]{0.937,0.937,0.937}}0.9200                       & {\cellcolor[rgb]{0.937,0.937,0.937}}0.9583                   & {\cellcolor[rgb]{0.937,0.937,0.937}}0.9451                   & {\cellcolor[rgb]{0.937,0.937,0.937}}0.9495               & {\cellcolor[rgb]{0.937,0.937,0.937}}0.9091 & {\cellcolor[rgb]{0.937,0.937,0.937}}0.9697               & {\cellcolor[rgb]{0.937,0.937,0.937}}0.9447 & {\cellcolor[rgb]{0.937,0.937,0.937}}0.040                     & {\cellcolor[rgb]{0.937,0.937,0.937}}0.040                        & {\cellcolor[rgb]{0.937,0.937,0.937}}0.000                    & {\cellcolor[rgb]{0.937,0.937,0.937}}0.020                    & {\cellcolor[rgb]{0.937,0.937,0.937}}0.050                & {\cellcolor[rgb]{0.937,0.937,0.937}}0.100 & {\cellcolor[rgb]{0.937,0.937,0.937}}0.010                & {\cellcolor[rgb]{0.937,0.937,0.937}}0.037 & {\cellcolor[rgb]{0.937,0.937,0.937}}0.9626     \\
                                             & Llama2-13b                                         & 1.0000                                                        & 0.9615                                                           & 0.9796                                                       & 0.9556                                                       & 0.9677                                                   & 0.9485                                     & 0.9592                                                   & 0.9674                                     & 0.000                                                         & 0.040                                                            & 0.000                                                        & 0.010                                                        & 0.010                                                    & 0.010                                     & 0.010                                                    & 0.011                                     & 0.9789                                         \\
                                             & {\cellcolor[rgb]{0.937,0.937,0.937}}Llama3-8b      & {\cellcolor[rgb]{0.937,0.937,0.937}}1.0000                    & {\cellcolor[rgb]{0.937,0.937,0.937}}0.9600                       & {\cellcolor[rgb]{0.937,0.937,0.937}}0.9583                   & {\cellcolor[rgb]{0.937,0.937,0.937}}0.9670                   & {\cellcolor[rgb]{0.937,0.937,0.937}}0.9783               & {\cellcolor[rgb]{0.937,0.937,0.937}}0.9697 & {\cellcolor[rgb]{0.937,0.937,0.937}}0.9216               & {\cellcolor[rgb]{0.937,0.937,0.937}}0.9650 & {\cellcolor[rgb]{0.937,0.937,0.937}}0.000                     & {\cellcolor[rgb]{0.937,0.937,0.937}}0.020                        & {\cellcolor[rgb]{0.937,0.937,0.937}}0.000                    & {\cellcolor[rgb]{0.937,0.937,0.937}}0.010                    & {\cellcolor[rgb]{0.937,0.937,0.937}}0.000                & {\cellcolor[rgb]{0.937,0.937,0.937}}0.010 & {\cellcolor[rgb]{0.937,0.937,0.937}}0.050                & {\cellcolor[rgb]{0.937,0.937,0.937}}0.013 & {\cellcolor[rgb]{0.937,0.937,0.937}}0.9770     \\
                                             & Llama3-70b                                         & 0.9901                                                        & 0.9515                                                           & 0.9796                                                       & 0.9677                                                       & 0.9787                                                   & 0.9800                                     & 0.9505                                                   & 0.9712                                     & 0.010                                                         & 0.040                                                            & 0.000                                                        & 0.020                                                        & 0.010                                                    & 0.010                                     & 0.030                                                    & 0.017                                     & 0.9808                                         \\
                                             & {\cellcolor[rgb]{0.937,0.937,0.937}}Vicuna-v1.5-7b & {\cellcolor[rgb]{0.937,0.937,0.937}}1.0000                    & {\cellcolor[rgb]{0.937,0.937,0.937}}0.9333                       & {\cellcolor[rgb]{0.937,0.937,0.937}}0.9697                   & {\cellcolor[rgb]{0.937,0.937,0.937}}0.9574                   & {\cellcolor[rgb]{0.937,0.937,0.937}}0.9892               & {\cellcolor[rgb]{0.937,0.937,0.937}}0.9691 & {\cellcolor[rgb]{0.937,0.937,0.937}}0.9278               & {\cellcolor[rgb]{0.937,0.937,0.937}}0.9638 & {\cellcolor[rgb]{0.937,0.937,0.937}}0.000                     & {\cellcolor[rgb]{0.937,0.937,0.937}}0.060                        & {\cellcolor[rgb]{0.937,0.937,0.937}}0.010                    & {\cellcolor[rgb]{0.937,0.937,0.937}}0.030                    & {\cellcolor[rgb]{0.937,0.937,0.937}}0.000                & {\cellcolor[rgb]{0.937,0.937,0.937}}0.000 & {\cellcolor[rgb]{0.937,0.937,0.937}}0.020                & {\cellcolor[rgb]{0.937,0.937,0.937}}0.017 & {\cellcolor[rgb]{0.937,0.937,0.937}}0.9761     \\
                                             & Vicuna-v1.5-13b                                    & 1.0000                                                        & 0.9231                                                           & 0.9796                                                       & 0.9677                                                       & 0.9892                                                   & 0.9703                                     & 0.9184                                                   & 0.9640                                     & 0.000                                                         & 0.060                                                            & 0.000                                                        & 0.020                                                        & 0.000                                                    & 0.020                                     & 0.030                                                    & 0.019                                     & 0.9761                                         \\
                                             & {\cellcolor[rgb]{0.937,0.937,0.937}}Gemma2-9b      & {\cellcolor[rgb]{0.937,0.937,0.937}}1.0000                    & {\cellcolor[rgb]{0.937,0.937,0.937}}0.9505                       & {\cellcolor[rgb]{0.937,0.937,0.937}}0.9796                   & {\cellcolor[rgb]{0.937,0.937,0.937}}0.9670                   & {\cellcolor[rgb]{0.937,0.937,0.937}}0.9684               & {\cellcolor[rgb]{0.937,0.937,0.937}}0.9608 & {\cellcolor[rgb]{0.937,0.937,0.937}}0.9505               & {\cellcolor[rgb]{0.937,0.937,0.937}}0.9681 & {\cellcolor[rgb]{0.937,0.937,0.937}}0.000                     & {\cellcolor[rgb]{0.937,0.937,0.937}}0.030                        & {\cellcolor[rgb]{0.937,0.937,0.937}}0.000                    & {\cellcolor[rgb]{0.937,0.937,0.937}}0.010                    & {\cellcolor[rgb]{0.937,0.937,0.937}}0.020                & {\cellcolor[rgb]{0.937,0.937,0.937}}0.030 & {\cellcolor[rgb]{0.937,0.937,0.937}}0.030                & {\cellcolor[rgb]{0.937,0.937,0.937}}0.017 & {\cellcolor[rgb]{0.937,0.937,0.937}}0.9789     \\ 
    \cline{2-19}
                                             & \textbf{Average}                                   & \textbf{0.9931}                                               & \textbf{0.9428}                                                  & \textbf{0.9721}                                              & \textbf{0.9611}                                              & \textbf{0.9744}                                          & \textbf{0.9582}                            & \textbf{0.9425}                                          & \textbf{0.9635}                            & \textbf{0.007}                                                & 0.041                                                            & \textbf{0.001}                                               & \textbf{0.017}                                               & \textbf{0.013}                                           & \textbf{0.026}                            & \textbf{0.026}                                           & \textbf{0.019}                            & \textbf{0.9758}                                \\ 
    \hline
    \rowcolor[rgb]{0.937,0.937,0.937} \multicolumn{2}{l||}{PlatonicDetector}                      & 0.9432                                                        & 0.8482                                                           & 0.9602                                                       & 0.8969                                                       & 0.8867                                                   & 0.8368                                     & 0.9182                                                   & 0.8986                                     & 0.066                                                         & 0.181                                                            & 0.021                                                        & 0.067                                                        & 0.104                                                    & 0.154                                     & 0.070                                                    & 0.095                                     & 0.9241                                         \\
    \multicolumn{2}{l||}{BD-LLM}                                                                  & 0.6944                                                        & 0.7299                                                           & 0.7619                                                       & 0.6452                                                       & 0.6087                                                   & 0.6545                                     & 0.6818                                                   & 0.6824                                     & 0.440                                                         & 0.410                                                            & 0.280                                                        & 0.380                                                        & 0.490                                                    & 0.250                                     & 0.370                                                    & 0.374                                     & 0.7226                                         \\
    \rowcolor[rgb]{0.937,0.937,0.937} \multicolumn{2}{l||}{OpenAIModerationAPI}                   & 0.1250                                                        & 0.2703                                                           & 0.6522                                                       & 0.6667                                                       & 0.8952                                                   & 0.8085                                     & 0.7010                                                   & 0.5884                                     & 0.100                                                         & 0.140                                                            & 0.120                                                        & 0.100                                                        & 0.110                                                    & 0.060                                     & 0.130                                                    & 0.109                                     & 0.7819                                         \\
    \multicolumn{2}{l||}{PerspectiveAPI}                                                          & 0.0377                                                        & 0.3030                                                           & 0.5476                                                       & 0.6494                                                       & 0.7957                                                   & 0.6947                                     & 0.6667                                                   & 0.5278                                     & 0.020                                                         & 0.060                                                            & 0.110                                                        & 0.060                                                        & 0.090                                                    & 0.120                                     & 0.100                                                    & 0.080                                     & 0.7704                                         \\
    \rowcolor[rgb]{0.937,0.937,0.937} \multicolumn{2}{l||}{PerplexityFilter}                      & 0.4767                                                        & 0.2678                                                           & 0.1739                                                       & 0.2457                                                       & 0.3196                                                   & 0.1973                                     & 0.1692                                                   & 0.2643                                     & 0.537                                                         & 0.500                                                            & 0.571                                                        & 0.474                                                        & 0.453                                                    & 0.449                                     & 0.501                                                    & 0.498                                     & 0.4472                                         \\
    \multicolumn{2}{l||}{WatchYourLanguage}                                                       & 0.3437                                                        & 0.2373                                                           & 0.9231                                                       & 0.7805                                                       & 0.8989                                                   & 0.6667                                     & 0.9109                                                   & 0.6801                                     & 0.030                                                         & \textbf{0.020}                                                   & 0.060                                                        & 0.040                                                        & 0.020                                                    & 0.060                                     & 0.050                                                    & 0.040                                     & 0.8479                                         \\
    \hline
    \end{tabular}%
    }
\end{table*}

\begin{table*}[h!]
    \caption{Evaluation Results on \newdataset{}.}
    \label{tab:new-data-set}
    \centering
    \resizebox{\textwidth}{!}{%
    \large

\begin{tabular}{l|l||cccccc|c||cccccc|c||c} 
    \hline
    \multicolumn{2}{l||}{\multirow{3}{*}{\textbf{\textbf{Detection Technique}}}}                  & \multicolumn{7}{c||}{\textbf{F1 Score}}                                                                                                                                                                                                                                                                                      & \multicolumn{7}{c||}{\textbf{False Positive Rate}}                                                                                                                                                                                                                                                                     & \multicolumn{1}{l}{\multirow{3}{*}{\textbf{Accuracy}}}  \\ 
    \cline{3-16}
    \multicolumn{2}{l||}{}                                                                        & \multicolumn{6}{c|}{Toxic Scenarios}                                                                                                                                                                                                                                        & \multicolumn{1}{l||}{\multirow{2}{*}{Average}} & \multicolumn{6}{c|}{Toxic Scenarios}                                                                                                                                                                                                                                  & \multicolumn{1}{l||}{\multirow{2}{*}{Average}} & \multicolumn{1}{l}{}                                    \\ 
    \cline{3-8}\cline{10-15}
    \multicolumn{2}{l||}{}                                                                        & Identity Attack                            & \multicolumn{1}{l}{Offense}                & \multicolumn{1}{l}{Flirtation}             & Profanity                                  & Sexually Explicit                          & Threat                                     & \multicolumn{1}{l||}{}                         & Identity Attack                           & \multicolumn{1}{l}{Offense}               & \multicolumn{1}{l}{Flirtation}            & Profanity                                 & Sexually Explicit                         & Threat                                    & \multicolumn{1}{l||}{}                         & \multicolumn{1}{l}{}                                    \\ 
    \hline
    \multirow{8}{*}{\rotatebox{90}{ToxicDetector}} & {\cellcolor[rgb]{0.937,0.937,0.937}}Llama2-7b      & {\cellcolor[rgb]{0.937,0.937,0.937}}0.9424 & {\cellcolor[rgb]{0.937,0.937,0.937}}0.8950 & {\cellcolor[rgb]{0.937,0.937,0.937}}0.8852 & {\cellcolor[rgb]{0.937,0.937,0.937}}0.9247 & {\cellcolor[rgb]{0.937,0.937,0.937}}0.9796 & {\cellcolor[rgb]{0.937,0.937,0.937}}0.9101 & {\cellcolor[rgb]{0.937,0.937,0.937}}0.9228     & {\cellcolor[rgb]{0.937,0.937,0.937}}0.010 & {\cellcolor[rgb]{0.937,0.937,0.937}}0.000 & {\cellcolor[rgb]{0.937,0.937,0.937}}0.020 & {\cellcolor[rgb]{0.937,0.937,0.937}}0.000 & {\cellcolor[rgb]{0.937,0.937,0.937}}0.000 & {\cellcolor[rgb]{0.937,0.937,0.937}}0.030 & {\cellcolor[rgb]{0.937,0.937,0.937}}0.010      & {\cellcolor[rgb]{0.937,0.937,0.937}}0.9283              \\
                                             & Llama2-13b                                         & 0.9749                                     & 0.9950                                     & 0.9694                                     & 0.9798                                     & 0.9849                                     & 0.9645                                     & 0.9781                                         & 0.020                                     & 0.010                                     & 0.010                                     & 0.010                                     & 0.010                                     & 0.020                                     & 0.013                                          & 0.9783                                                  \\
                                             & {\cellcolor[rgb]{0.937,0.937,0.937}}Llama3-8b      & {\cellcolor[rgb]{0.937,0.937,0.937}}0.9950 & {\cellcolor[rgb]{0.937,0.937,0.937}}0.9849 & {\cellcolor[rgb]{0.937,0.937,0.937}}0.9588 & {\cellcolor[rgb]{0.937,0.937,0.937}}0.9424 & {\cellcolor[rgb]{0.937,0.937,0.937}}0.9412 & {\cellcolor[rgb]{0.937,0.937,0.937}}0.9697 & {\cellcolor[rgb]{0.937,0.937,0.937}}0.9653     & {\cellcolor[rgb]{0.937,0.937,0.937}}0.000 & {\cellcolor[rgb]{0.937,0.937,0.937}}0.010 & {\cellcolor[rgb]{0.937,0.937,0.937}}0.010 & {\cellcolor[rgb]{0.937,0.937,0.937}}0.010 & {\cellcolor[rgb]{0.937,0.937,0.937}}0.080 & {\cellcolor[rgb]{0.937,0.937,0.937}}0.020 & {\cellcolor[rgb]{0.937,0.937,0.937}}0.022      & {\cellcolor[rgb]{0.937,0.937,0.937}}0.9658              \\
                                             & Llama3-70b                                         & 0.9950                                     & 0.9849                                     & 0.9798                                     & 0.9950                                     & 0.9751                                     & 0.9950                                     & 0.9875                                         & 0.010                                     & 0.010                                     & 0.010                                     & 0.000                                     & 0.030                                     & 0.010                                     & 0.012                                          & 0.9875                                                  \\
                                             & {\cellcolor[rgb]{0.937,0.937,0.937}}Vicuna-v1.5-7b & {\cellcolor[rgb]{0.937,0.937,0.937}}0.9447 & {\cellcolor[rgb]{0.937,0.937,0.937}}0.9583 & {\cellcolor[rgb]{0.937,0.937,0.937}}0.9479 & {\cellcolor[rgb]{0.937,0.937,0.937}}0.9278 & {\cellcolor[rgb]{0.937,0.937,0.937}}0.9412 & {\cellcolor[rgb]{0.937,0.937,0.937}}0.9300 & {\cellcolor[rgb]{0.937,0.937,0.937}}0.9417     & {\cellcolor[rgb]{0.937,0.937,0.937}}0.050 & {\cellcolor[rgb]{0.937,0.937,0.937}}0.000 & {\cellcolor[rgb]{0.937,0.937,0.937}}0.010 & {\cellcolor[rgb]{0.937,0.937,0.937}}0.040 & {\cellcolor[rgb]{0.937,0.937,0.937}}0.080 & {\cellcolor[rgb]{0.937,0.937,0.937}}0.070 & {\cellcolor[rgb]{0.937,0.937,0.937}}0.042      & {\cellcolor[rgb]{0.937,0.937,0.937}}0.9425              \\
                                             & Vicuna-v1.5-13b                                    & 0.9749                                     & 0.9950                                     & 0.9798                                     & 0.9798                                     & 0.9697                                     & 0.9802                                     & 0.9799                                         & 0.020                                     & 0.010                                     & 0.010                                     & 0.010                                     & 0.020                                     & 0.030                                     & 0.017                                          & 0.9800                                                  \\
                                             & {\cellcolor[rgb]{0.937,0.937,0.937}}Gemma2-9b      & {\cellcolor[rgb]{0.937,0.937,0.937}}0.9900 & {\cellcolor[rgb]{0.937,0.937,0.937}}0.9849 & {\cellcolor[rgb]{0.937,0.937,0.937}}0.9800 & {\cellcolor[rgb]{0.937,0.937,0.937}}0.9198 & {\cellcolor[rgb]{0.937,0.937,0.937}}0.9412 & {\cellcolor[rgb]{0.937,0.937,0.937}}0.9697 & {\cellcolor[rgb]{0.937,0.937,0.937}}0.9643     & {\cellcolor[rgb]{0.937,0.937,0.937}}0.010 & {\cellcolor[rgb]{0.937,0.937,0.937}}0.010 & {\cellcolor[rgb]{0.937,0.937,0.937}}0.020 & {\cellcolor[rgb]{0.937,0.937,0.937}}0.010 & {\cellcolor[rgb]{0.937,0.937,0.937}}0.080 & {\cellcolor[rgb]{0.937,0.937,0.937}}0.020 & {\cellcolor[rgb]{0.937,0.937,0.937}}0.025      & {\cellcolor[rgb]{0.937,0.937,0.937}}0.9650              \\ 
    \cline{2-17}
                                             & \textbf{Average}                                   & \textbf{0.9738}                            & \textbf{0.9712}                            & \textbf{0.9573}                            & \textbf{0.9528}                            & \textbf{0.9618}                            & \textbf{0.9599}                            & \textbf{0.9628}                                & \textbf{0.017}                            & \textbf{0.007}                            & \textbf{0.013}                            & \textbf{0.011}                            & 0.043                                     & 0.029                                     & \textbf{0.020}                                 & \textbf{0.9639}                                         \\ 
    \hline
    \rowcolor[rgb]{0.937,0.937,0.937} \multicolumn{2}{l||}{PlatonicDetector}                      & 0.9166                                     & 0.9132                                     & 0.4943                                     & 0.8901                                     & 0.8500                                     & 0.9259                                     & 0.8317                                         & 0.189                                     & 0.179                                     & 0.369                                     & 0.257                                     & 0.203                                     & 0.104                                     & 0.217                                          & 0.8357                                                  \\
    \multicolumn{2}{l||}{BD-LLM}                                                                  & 0.7800                                     & 0.6211                                     & 0.8454                                     & 0.8223                                     & 0.7826                                     & 0.8316                                     & 0.7805                                         & 0.220                                     & 0.110                                     & 0.120                                     & 0.160                                     & 0.120                                     & 0.110                                     & 0.140                                          & 0.7983                                                  \\
    \rowcolor[rgb]{0.937,0.937,0.937} \multicolumn{2}{l||}{OpenAIModerationAPI}                   & 0.9238                                     & 0.8776                                     & 0.7709                                     & 0.8856                                     & 0.9320                                     & 0.9293                                     & 0.8865                                         & 0.130                                     & 0.100                                     & 0.100                                     & 0.120                                     & 0.100                                     & 0.060                                     & 0.102                                          & 0.8900                                                  \\
    \multicolumn{2}{l||}{PerspectiveAPI}                                                          & 0.9378                                     & 0.9608                                     & 0.5455                                     & 0.9346                                     & 0.9565                                     & 0.8691                                     & 0.8674                                         & 0.110                                     & 0.060                                     & 0.040                                     & 0.140                                     & 0.080                                     & 0.080                                     & 0.085                                          & 0.8883                                                  \\
    \rowcolor[rgb]{0.937,0.937,0.937} \multicolumn{2}{l||}{PerplexityFilter}                      & 0.4672                                     & 0.4616                                     & 0.5056                                     & 0.4670                                     & 0.4897                                     & 0.4726                                     & 0.4773                                         & 0.527                                     & 0.443                                     & 0.530                                     & 0.581                                     & 0.481                                     & 0.504                                     & 0.511                                          & 0.4898                                                  \\
    \multicolumn{2}{l||}{WatchYourLanguage}                                                       & 0.8667                                     & 0.8070                                     & 0.4706                                     & 0.9158                                     & 0.7836                                     & 0.7607                                     & 0.7674                                         & 0.020                                     & 0.020                                     & 0.040                                     & 0.030                                     & \textbf{0.040}                            & \textbf{0.010}                            & 0.027                                          & 0.8158                                                  \\
    \hline
    \end{tabular}%
    }
\end{table*}

\noindent\textbf{LLMs under test.} We select \revision{seven} popular open-source LLMs for our evaluation. These include various versions of the LLama models, chosen for their widespread use and robust performance in natural language processing tasks. The specific models tested are:

\begin{itemize}
    \item \textbf{LLama-3 (8B and 70B versions)}~\cite{llama3}: These models are the latest iterations in the LLama series, offering significant improvements in both size and performance. The 8 billion parameter model (8B) and the 70 billion parameter model (70B) are tested to evaluate performance across different scales.
    
    \item \textbf{LLama-2 (7B and 13B versions)}~\cite{llama2}: As the second generation of LLama models, these versions provide enhancements in efficiency and accuracy. 
    
    \item \textbf{LLama (7B and 13B versions)}~\cite{llama}: We use Vicuna-v1.5-7B~\cite{vicuna} and Vicuna-v1.5-13B~\cite{vicuna13} which are fine-tuning from origin LLama.
    \item \revision{\textbf{Gemma-2 (9B)}~\cite{gemmateam2024gemma2improvingopen}: We select this latest LLM developed by Google to evaluate the generalization of \tool{} across different models.}
\end{itemize}

\noindent\textbf{Metrics.} To evaluate the effectiveness of \tool{}, we employ the following metrics:

\begin{itemize}
    \item \textbf{F1 Score}: This metric provides a balance between precision and recall, giving a single measure of a test's accuracy. It is especially useful when the class distribution is imbalanced.
    \item \textbf{False Positive Rate}: This metric measures the proportion of benign prompts incorrectly classified as toxic. A lower FPR indicates fewer false alarms.
    \item \textbf{Accuracy}: This metric represents the proportion of correctly classified prompts (both benign and toxic) out of all prompts. It gives a general sense of the model's overall performance.
    \item \textbf{ROC Curve}: The Receiver Operating Characteristic (ROC) curve illustrates the true positive rate (sensitivity) against the false positive rate. This curve helps visualize the trade-offs between true positives and false positives and is useful for comparing different models.
\end{itemize}

\noindent\textbf{Experimental Settings.} All experiments are conducted using two Titan RTX 48 GPUs on Ubuntu 22.04. We configure all baselines and LLMs according to their respective instructions. To train the classifier, we use a fully connected MLP with 5 layers and 300 million parameters. The training parameters are as follows: a batch size of 20, 100 epochs, a learning rate of 0.01, and a weight decay of 0.0002. For the training dataset, we use 50\% of the data for training and the rest for testing. To mitigate the effects of randomness in evaluation, we ran all experiments ten times.

\begin{figure*}[h!t!]
    \centering
    \includegraphics[width=0.88\linewidth]{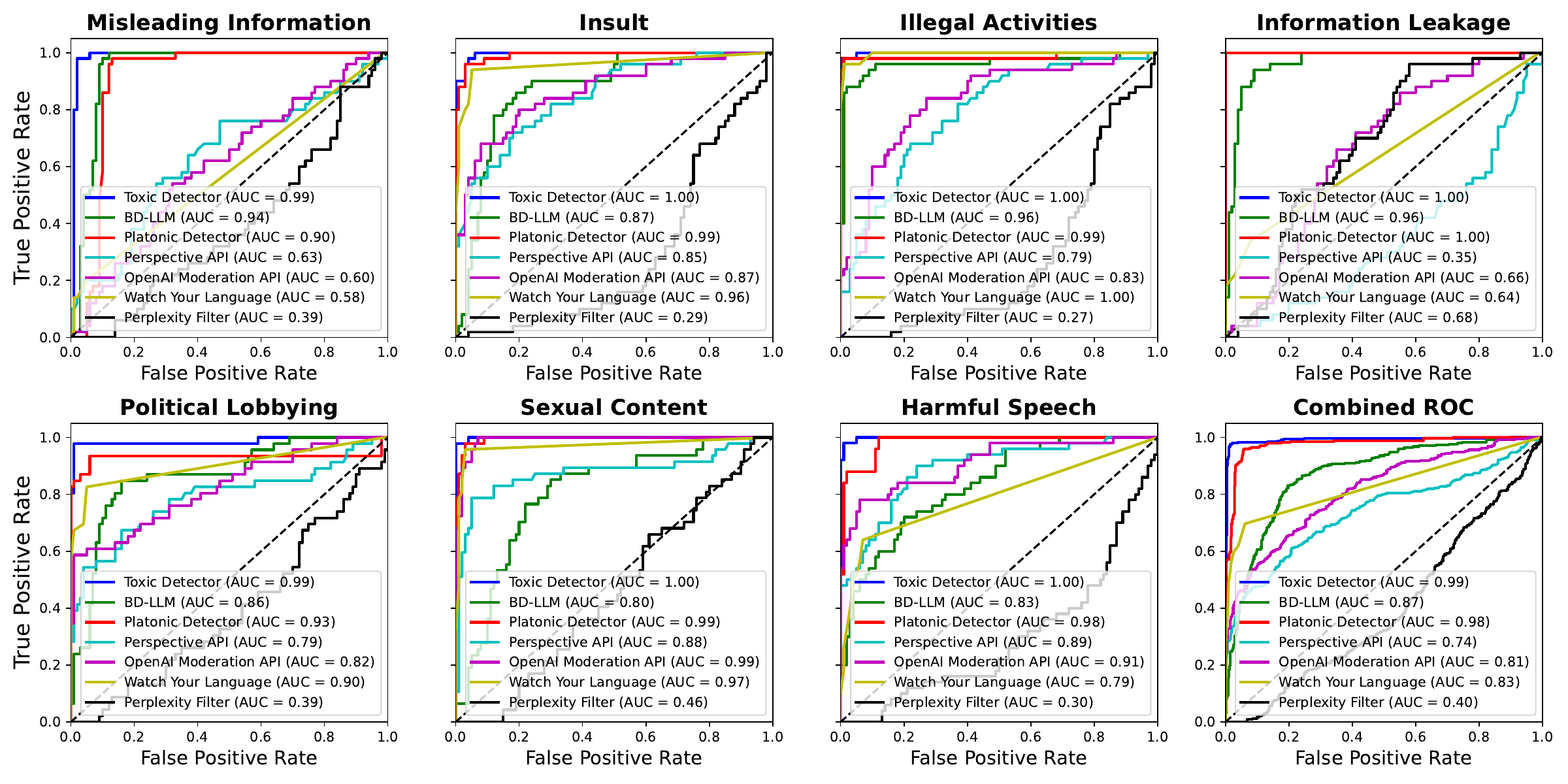}
    \caption{The ROC curves for identifying seven different types of toxic scenarios, comparing the performance of \tool{} on all LLMs under test with all baselines (\ourdataset{}).}
    \label{fig:roc}
\end{figure*}

\subsection{RQ1 (Effectiveness)}
\label{sec:rq1-effectiveness}

In this research question, we aim to evaluate the effectiveness of \tool{} in accurately identifying toxic prompts across various scenarios. We compare \tool{} with other baseline methods across multiple LLMs under test. \revision{The results are summarized in Table~\ref{tab:our-dataset}, Table~\ref{tab:new-data-set}, Figure~\ref{fig:roc}, and Table~\ref{tab:f1_scores}.~\footnote{\revision{Due to page limitation, we only present the detailed data of \tool{} and leave other baselines' details on our website~\cite{ToxicDet:online}.}}}

\revision{\textbf{Comparison between Different Models.} Table~\ref{tab:our-dataset} presents the average F1 scores, false positive rates, and overall accuracies of various classifiers in identifying toxic prompts across different scenarios (statistically significant results are highlighted in bold, calculated using the Mann-Whitney U test~\cite{mcknight2010mann} at a 0.05 confidence level). The results indicate that methods like \textsc{PerplexityFilter} and \textsc{BD-LLM} struggle with high false positive rates, reflecting difficulties in accurately distinguishing between toxic and benign prompts. For example, \textsc{PerplexityFilter} has a false positive rate of 0.498, leading to numerous false alarms. In contrast, \tool{} achieves a low false positive rate of 0.019, demonstrating its precision in differentiating between toxic and benign prompts—a crucial quality for practical applications where avoiding unnecessary disruptions is critical. Furthermore, \tool{} achieves the highest average F1 score, 0.9635, across both Gemma-2 and Llama series LLMs, underscoring its robust capability in detecting toxic prompts. The superior performance of \tool{} can be attributed to its efficient use of embedding vectors and a lightweight MLP classifier, which together enhance its detection capabilities.}

\revision{An interesting finding is that models with larger parameter sizes (e.g., Llama2-13b vs. Llama2-7b) and newer architectures (e.g., Llama3 vs. Llama2) are more effective in refusing toxic prompts in their responses, benefiting from sophisticated alignment techniques~\cite{llama2,llama3}. Additionally, \tool{} shows better results with larger and newer models, providing evidence that these models are trained with better semantic embeddings that can represent high-level concepts, including toxic ones.}

\revision{\textbf{Comparison between Different Datasets.} In Table~\ref{tab:new-data-set}, we evaluate all baselines on an orthogonal dataset, \newdataset{}. \tool{} once again achieves the best performance, with an average F1 score of 0.9628 and an exceptionally low false positive rate of 0.02. Methods relying on pre-trained models, such as \textsc{WatchYourLanguage}, \textsc{PerspectiveAPI}, and \textsc{OpenAIModerationAPI}, show significant increases in average F1 scores (from 0.6801 to 0.7674, 0.5278 to 0.8674, and 0.5884 to 0.8865, respectively), likely because their pre-training data includes \newdataset{} (created in 2020). Conversely, \textsc{PlatonicDetector}'s performance drops significantly, with its average F1 score falling to 0.8317 and its false positive rate increasing to 0.217, indicating a lack of generalization to different distributions of toxic prompts. These results demonstrate \tool{}'s ability to maintain robust performance across different datasets and toxic scenarios, consistently outperforming a range of baseline methods in both precision and generalizability.}

\revision{\textbf{Comparison with \textsc{PlatonicDetector}.} While \tool{} draws inspiration from the feature construction approach used in \textsc{PlatonicDetector}, we have implemented significant improvements. The original work~\cite{huh2024platonic} demonstrates that different LLMs produce consistent embeddings for the same concepts by analyzing the similarity of the last token's embedding. \tool{} extends this concept into a robust pipeline specifically designed to detect toxic prompts. Specifically, \tool{} concatenates embeddings from each transformer layer, rather than relying solely on the last layer as \textsc{PlatonicDetector} does. This method allows \tool{} to capture a more comprehensive set of features, significantly improving its ability to identify toxic prompts. Our comparative results highlight this enhancement, with \tool{} achieving statistically significantly better results across most toxic scenarios on both \ourdataset{} and \newdataset{}, in terms of F1 score and false positive rate (as determined by the Mann-Whitney U test at a 0.05 confidence level). Additionally, we have translated the theoretical insights from \textsc{PlatonicDetector} into a practical, automated pipeline, integrating LLM-based data augmentation. This provides developers with a powerful tool to leverage LLMs more effectively in their applications, thereby enhancing software safety and trustworthiness.}

\revision{\textbf{Comparison in terms of ROC.}} Figure~\ref{fig:roc} presents the averaged ROC curves for identifying seven different types of toxic scenarios across various LLMs tested on \ourdataset{} \revision{(detailed results for \newdataset{} are available on our website~\cite{ToxicDet:online})}. The figure shows that methods like \textsc{PerspectiveAPI} and \textsc{PerplexityFilter} have lower Area Under the Curve (AUC) values, indicating less reliable detection capabilities. For example, \textsc{PerplexityFilter} has an AUC of 0.35 in the ``Information Leakage'' scenario, reflecting poor performance. In contrast, \tool{} consistently achieves higher AUC values across all types of malicious activities, with an overall AUC of 0.99 on both \ourdataset{} and \newdataset{}. These high AUC values demonstrate \tool{}'s robustness in detecting toxic content across diverse scenarios. \tool{}'s effectiveness is further underscored by its ability to maintain high accuracy and reliability, even in complex and varied contexts.

\begin{table}[t!]
    \caption{F1 Scores for different toxic scenarios with jailbreaking on \ourdataset{} and \newdataset{}.}
    \label{tab:toxic_scenarios_with_jailbreak}
    \centering
    \resizebox{\linewidth}{!}{%
    \large
    \begin{tabular}{l|c|c} 
    \hline
    \multirow{2}{*}{\textbf{Model}}                  & \multicolumn{2}{c}{\textbf{Dataset}}           \\ 
    \cline{2-3}
                                                     & \ourdataset{} & \newdataset{}  \\ 
    \hline
    \rowcolor[rgb]{0.937,0.937,0.937} Llama2-7b      & 0.9463                  & 0.9749               \\
    Llama2-13b                                       & 0.9706                  & 0.9577               \\
    \rowcolor[rgb]{0.937,0.937,0.937} Llama3-8b      & 0.9365                  & 0.9664               \\
    Llama3-70b                                       & 0.9951                  & 0.9799               \\
    \rowcolor[rgb]{0.937,0.937,0.937} Vicuna-v1.5-7b & 0.9299                  & 0.9778               \\
    Vicuna-v1.5-13b                                  & 0.9344                  & 0.9483               \\
    \rowcolor[rgb]{0.937,0.937,0.937} Gemma2-9b      & 0.9521                  & 0.9434               \\ 
    \hline
    \textbf{Average}                                 & \textbf{0.9521}         & \textbf{0.9641}      \\
    \hline
    \end{tabular}
    }
\end{table}

\begin{table}[t!]
    \caption{Comparison of F1 Scores for different toxic scenarios with and without concept prompt augmentation, and the corresponding boost on \ourdataset{}. Values in bold indicate the highest F1 Score in each scenario. }
    \label{tab:f1_scores}
    \centering
    \resizebox{\linewidth}{!}{%
    \large
    \begin{tabular}{c||ccc}
        \hline
        \rowcolor[HTML]{FFFFFF}
        \multicolumn{1}{c||}{\textbf{Toxic Scenario}} & \textbf{F1 Score (Plain)} & \textbf{F1 Score (Aug)} & \textbf{Boost} \\
        \hline
        \rowcolor[HTML]{EFEFEF}
        Information Leakage & 0.9434 & \textbf{1.0000} & 0.0566 \\
        \rowcolor[HTML]{FFFFFF}
        Misleading Information & 0.8958 & \textbf{0.9200} & 0.0242 \\
        \rowcolor[HTML]{EFEFEF}
        Illegal Activities & 0.9697 & \textbf{0.9800} & 0.0103 \\
        \rowcolor[HTML]{FFFFFF}
        Political Lobbying & \textbf{0.9451} & \textbf{0.9451} & - \\
        \rowcolor[HTML]{EFEFEF}
        Sexual Content & 0.9495 & \textbf{0.9583} & 0.0088 \\
        \rowcolor[HTML]{FFFFFF}
        Harmful Speech & 0.8913 & \textbf{0.9167} & 0.0254 \\
        \rowcolor[HTML]{EFEFEF}
        Insult & 0.8140 & \textbf{0.9697} & 0.1557 \\
        \hline
        \hline
        \rowcolor[HTML]{FFFFFF}
        Overall & 0.9155 & \textbf{0.9557} & 0.0401 \\
        \hline
    \end{tabular}
    }
\end{table}

\begin{table}[t!]
    \caption{F1 Score Across Varying Similarity Thresholds on \ourdataset{}}
    \label{tab:similarity_f1}
    \centering
    \resizebox{\linewidth}{!}{%
    \large

\begin{tblr}{
  column{2} = {c},
  column{3} = {c},
  column{4} = {c},
  column{5} = {c},
  column{6} = {c},
  column{7} = {c},
  cell{1}{1} = {r=2}{},
  cell{1}{2} = {c=10}{},
  hline{1,3-4} = {-}{},
}
\textbf{Metric} & \textbf{Similarity Threshold} &        &        &        &        &      &        &        &        &        \\
                & 0.1                           & 0.2    & 0.3    & 0.4    & 0.5    & 0.6  & 0.7    & 0.8    & 0.9    & 1.0    \\
F1 Score        & 0.5939                        & 0.5786 & 0.7907 & 0.9412 & 0.9505 & 0.9600 & 0.9505 & 0.9703 & 0.9703 & 0.9293 
\end{tblr}%
    }
\end{table}

\revision{\textbf{Toxic Detection with Jailbreak Techniques.} Table~\ref{tab:toxic_scenarios_with_jailbreak} presents the average results of \tool{} in detecting toxic prompts using template-based jailbreak techniques, measured by F1 score. We populate manually crafted jailbreak templates from previous work~\cite{liu2023jailbreaking} with toxic prompts as input for \tool{}. The results show that \tool{} achieves an average F1 score of 0.9521 on \ourdataset{} and 0.9641 on \newdataset{}, demonstrating that \tool{} effectively identifies toxic prompts embedded within jailbreak techniques, even when trained on plain toxic prompts.}

\revision{\textbf{Effectiveness of Concept Prompt Augmentation.}} Table~\ref{tab:f1_scores} compares the F1 scores for different toxic scenarios with and without concept prompt augmentation, along with the corresponding performance boost on \ourdataset{}. \tool{} shows significant improvement with concept prompt augmentation, particularly in scenarios like ``Insult,'' where the F1 score jumps from 0.8140 to 0.9697. This notable boost demonstrates the added value of concept prompt augmentation in enhancing detection accuracy. The ability of \tool{} to leverage these augmentations effectively highlights its superior design and implementation.

\revision{\textbf{Validation of Concept Prompts.} To validate the generated toxic concept prompts, three authors independently review 680 of these prompts. The goal is to perform a binary classification to determine whether each generated toxic concept prompt falls within the disallowed usages outlined by OpenAI~\cite{Usagepol82:online}. This is important because OpenAI, with the largest user base~\cite{Numberof91:online} and government regulation~\cite{FACTSHEE29:online}, provides a representative taxonomy of toxic prompts. The results show that all three authors unanimously agree that the generated concept prompts align with the disallowed usage criteria. This consensus demonstrates the effectiveness of the concept prompt augmentation process.}

\revision{\textbf{Similarity Threshold.} We empirically decide the similarity threshold when selecting toxic concept prompts by analyzing the F1 score across varying thresholds on \ourdataset{}. As shown in Table~\ref{tab:similarity_f1}, the results show that the F1 score increases as the similarity threshold rises, peaking between 0.8 and 0.9, and then slightly decreases at 1.0. We set a high similarity threshold to focus on toxic prompts while still maintaining diversity among toxic concept prompts. For example, this approach allows us to capture similar concepts with diverse toxic prompts, such as ``Inquire about private access credentials,'' ``Seek disclosure of proprietary data,'' and ``Ask for unreleased financial reports.''}

The evaluation results clearly demonstrate that \tool{} outperforms existing methods in detecting toxic prompts across a wide range of scenarios. The consistently high F1 scores and low false positive rates indicate that \tool{} is both accurate and reliable. The robustness of \tool{} is further evidenced by its high AUC values across different types of toxic content, showing its superior performance compared to other classifiers. Concept prompt augmentation significantly enhances detection effectiveness, as shown by the improvements in F1 scores. These findings imply the practical utility of \tool{} in real-world applications.

\begin{tcolorbox}[title=Answer to RQ1,boxrule=1pt,boxsep=1pt,left=2pt,right=2pt,top=2pt,bottom=2pt]
\tool{} demonstrates superior performance in detecting toxic prompts across various LLMs compared to existing methods. Concept prompt augmentation significantly enhances detection effectiveness.
\end{tcolorbox}

\subsection{RQ2 (Efficiency)}
\label{sec:rq2-efficiency}

\begin{figure}[t!]
    \centering
    \includegraphics[width=\linewidth]{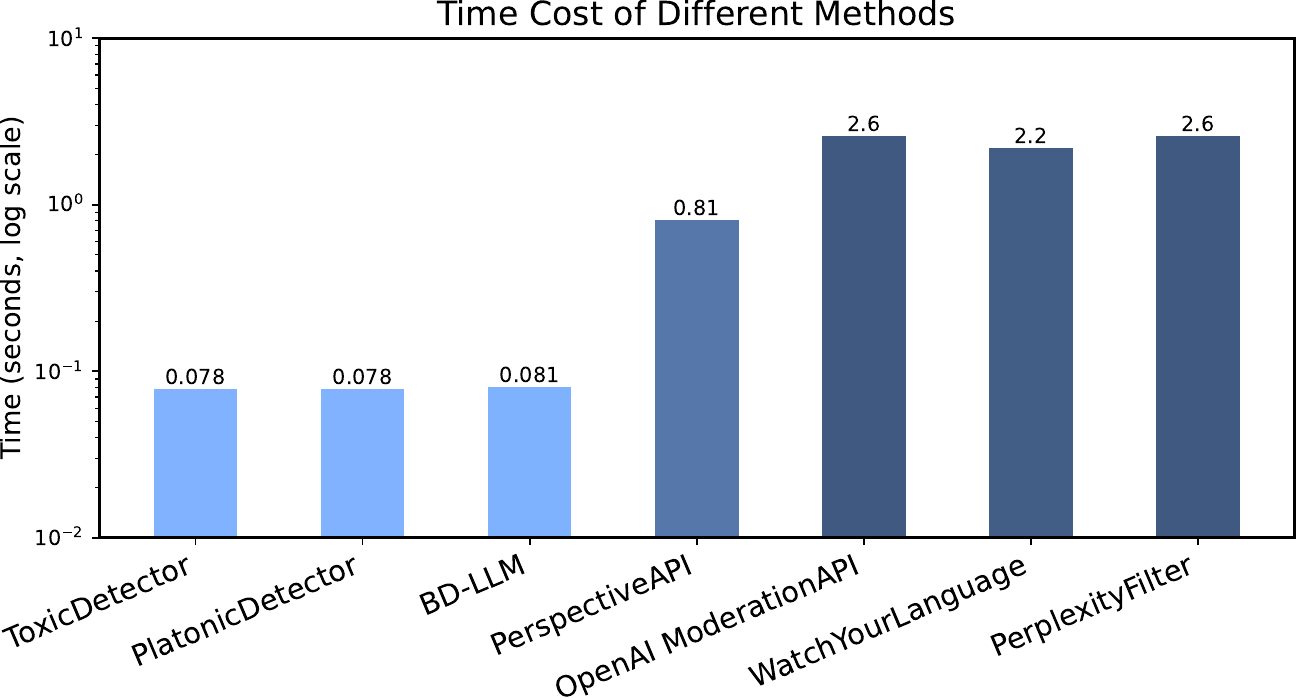}
    \caption{Comparison of the average prompt processing times in our evaluation, displayed on a logarithmic scale.}
    \label{fig:time_cost}
\end{figure}

In this research question, we aim to evaluate the training efforts and inference time cost of \tool{}. We train \tool{} with different training set sizes and record the training time. Additionally, we measure the classification time during toxic detection at runtime. Table~\ref{tab:train_times} and Figure~\ref{fig:time_cost} summarize the results.

Table~\ref{tab:train_times} illustrates the relationship between the number of training epochs, the corresponding training times, and the resulting F1 scores for our model. As the number of training epochs increases from 20 to 200, the training time also increases, starting at 69.4 seconds and reaching 197.6 seconds. With the increase in training time, the F1 scores show significant improvement, beginning at 0.942 with 20 epochs and peaking at 0.980 at 100 epochs. Beyond 100 epochs, the F1 score stabilizes at 0.980, indicating that additional training does not further enhance the model's performance. This table highlights the balance between training duration and model accuracy, suggesting that 100 epochs is optimal for achieving high performance without unnecessary extra training time. Additionally, the relatively short training times, even at maximum epochs, demonstrate that \tool{} is fast to train.

\begin{table}[t!]
    \caption{Training epochs, corresponding training times, and F1 scores. }
    \label{tab:train_times}
    \centering
    \resizebox{\linewidth}{!}{%
    \large
    \begin{tabular}{lcccccc}
        \hline
        \rowcolor[HTML]{FFFFFF} 
        \cellcolor[HTML]{FFFFFF}                                         & \multicolumn{6}{c}{\cellcolor[HTML]{FFFFFF}\textbf{Train Epochs}} \\
        \rowcolor[HTML]{FFFFFF} 
        \multirow{-2}{*}{\cellcolor[HTML]{FFFFFF}\textbf{Metric}} & 20 & 40 & 60 & 80 & 100 & 200 \\ \hline
        \rowcolor[HTML]{EFEFEF} 
        Train Time (seconds) & 69.4 & 75.8 & 79.2 & 88.2 & 89.4 & 197.6 \\
        \rowcolor[HTML]{FFFFFF} 
        F1 Score & 0.942 & 0.960 & 0.970 & 0.951 & 0.980 & 0.980 \\
        \hline
    \end{tabular}%
    }
\end{table}

\revision{Figure~\ref{fig:time_cost} compares the average prompt processing times for different methods. Methods like \textsc{PerplexityFilter}, \textsc{WatchYourLanguage}, and \textsc{OpenAIModerationAPI} show longer processing times, ranging from 2.2 to 2.6 seconds, reflecting computational overhead or network latency. In contrast, smaller models demonstrate remarkable efficiency, with \textsc{BD-LLM} processing a prompt in 0.081 seconds, and \textsc{ToxicDetector} and \textsc{PlatonicDetector} achieving the lowest processing times of approximately 0.078 seconds. The low processing time of \textsc{ToxicDetector} indicates its suitability for real-time applications, making it highly efficient in environments where prompt response times are critical. This efficiency can be attributed to \textsc{ToxicDetector}'s streamlined feature extraction and lightweight MLP classifier.}

\begin{figure}[t!]
    \centering
    \begin{subfigure}[b]{0.22\textwidth}
        \centering
        \includegraphics[width=\textwidth]{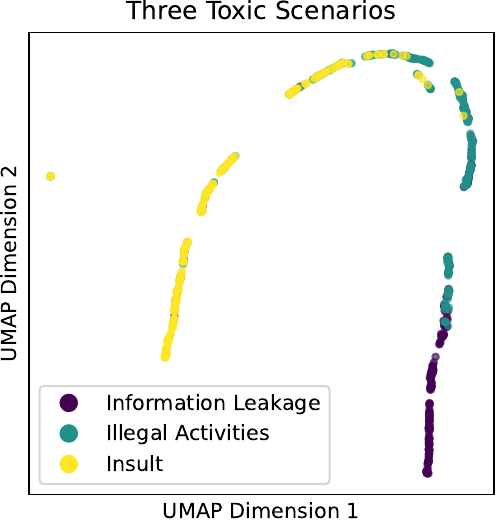}
        \caption{Different toxic scenarios.}
        \label{fig:umap_plot3type}
    \end{subfigure}
    \hfill
    \begin{subfigure}[b]{0.22\textwidth}
        \centering
        \includegraphics[width=\textwidth]{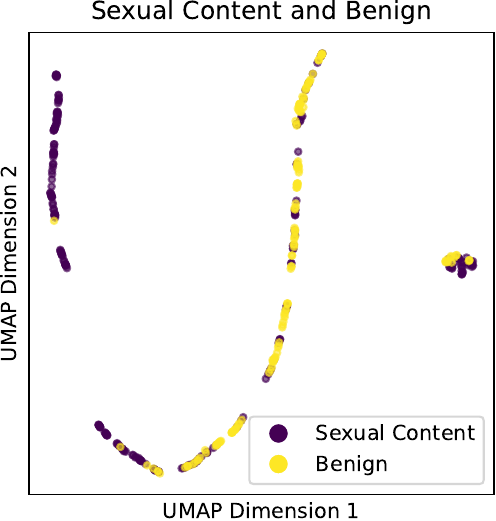}
        \caption{Toxic \& benign prompts. }
        \label{fig:umap_plot_moderation}
    \end{subfigure}
    \caption{Visualization of Prompt Embeddings by UMAP~\cite{mcinnes2020umap}.}
\label{fig:umap_plot}
\end{figure}

Training efforts and inference time are critical factors in the deployment of machine learning models, especially in real-time applications. Efficient training processes allow for quicker updates and retraining cycles, ensuring that models can adapt to new data and evolving scenarios without significant downtime. Short inference times are equally important as they enable the model to provide rapid responses, which is crucial in applications such as content moderation, online safety, and customer service. High training and inference efficiency also reduce computational resource consumption, making the system more cost-effective and scalable. Overall, optimizing both training efforts and inference time enhances the practicality and responsiveness of the deployed model.

\begin{tcolorbox}[title=Answer to RQ2,boxrule=1pt,boxsep=1pt,left=2pt,right=2pt,top=2pt,bottom=2pt]
\tool{} requires minimal training efforts and achieves fast inference times for detecting toxic prompts, which is crucial for real-world applications.
\end{tcolorbox}

\subsection{RQ3 (Feature Representation)}

We qualitatively examine why the feature representation effectively identifies toxic prompts using UMAP~\cite{mcinnes2020umap} for dimensionality reduction, as shown in Figure~\ref{fig:umap_plot}.

Figure~\ref{fig:umap_plot3type} displays UMAP results for three toxic scenarios—Information Leakage, Illegal Activities, and Insult—on LLama-2 7B~\cite{llama2}. The prompts form distinct clusters, indicating that the feature representation accurately captures the unique characteristics of each toxic type. This separation demonstrates the method's ability to differentiate between various toxic prompts reliably.

Figure~\ref{fig:umap_plot_moderation} contrasts toxic and benign prompts on LLama-2 7B~\cite{llama2}. The clear distinction between sexual content and benign clusters confirms that the feature representation effectively distinguishes toxic from non-toxic prompts, reducing false positives.

Overall, the UMAP visualizations confirm that \tool{}'s feature representation robustly differentiates between multiple toxic scenarios and benign prompts, ensuring accurate and reliable toxic prompt detection.

\begin{tcolorbox}[title=Answer to RQ3,boxrule=1pt,boxsep=1pt,left=2pt,right=2pt,top=2pt,bottom=2pt]
The feature representation clearly distinguishes different toxic scenarios and separates toxic from benign prompts, enabling effective and accurate detection of toxic prompts by classifiers.
\end{tcolorbox}

\vspace{-1em}
\section{Threats To Validity}

Our evaluation of \tool{} faces several potential threats. 

First, dataset construction may introduce biases. Although we combined multiple benchmarks to create a comprehensive dataset, the selected toxic and benign prompts might not fully capture the diversity of real-world interactions. Additionally, benign prompts from ShareGPT may not represent typical user behavior across all platforms, potentially limiting the generalizability of our results.

Second, experimental settings such as the choice of LLMs and baseline configurations could affect our findings. We used popular open-source LLMs configured per their guidelines, but different model versions or implementations might yield varying performance. Moreover, \tool{}'s effectiveness could differ when applied to other LLMs or in different application contexts.

Lastly, the training and evaluation process itself may pose validity threats. Despite conducting multiple runs and using standard metrics to reduce randomness, inherent variability in machine learning experiments can cause slight result fluctuations. The classifier's hyperparameters, like learning rate and batch size, were selected based on preliminary tests and might not be optimal universally. Future research should investigate these parameters further to enhance \tool{}'s robustness.

\vspace{-0.5em}
\section{Related Work}
\label{sec:related}

\vspace{-0.5em}
\subsection{Toxic Prompts}
Toxic prompts are inputs that cause LLMs to generate harmful or inappropriate responses, making their detection crucial for safe interactions. Datasets like RealToxicityPrompts~\cite{gehman2020realtoxicityprompts} provide benchmarks to assess LLMs' tendency to produce toxic content, underscoring the importance of robust detection mechanisms for responsible language model deployment.

\vspace{-0.6em}
\subsection{Jailbreaking on LLMs}
LLMs are susceptible to jailbreak attacks that leverage toxic prompts to produce unethical outputs. Studies such as Liu et al.~\cite{liu2023jailbreaking} and MASTERKEY~\cite{Deng_2024} demonstrate how adversarial prompts can bypass safeguards in models like \textsc{ChatGPT} and \textsc{Bard}. These vulnerabilities highlight the need for effective defenses, which \tool{} addresses by detecting and mitigating toxic prompts before they exploit these weaknesses.

\vspace{-0.6em}
\subsection{Toxic Prompt Detection Methods}
Detecting toxic prompts is crucial for the safe and ethical deployment of LLMs. Various methods have been proposed to identify and mitigate the effects of toxic prompts.

\textbf{Whitebox Methods:} These methods leverage the internal state of the model to detect toxic content. For instance, \textsc{PlatonicDetector}~\cite{huh2024platonic} utilizes the convergent representations in LLMs to identify toxic prompts, offering insights into the underlying dynamics of language processing. \textsc{PerplexityFilter}~\cite{jain2023baseline}, on the other hand, assesses the model's confidence in its responses, filtering out prompts that elicit low-confidence responses as potentially toxic. This approach is particularly effective in isolating subtle or cleverly disguised toxic content that may not trigger traditional detection mechanisms.

\textbf{Blackbox Methods:} These methods use pre-trained models without accessing their internal states. The \textsc{OpenAI Moderation API}~\cite{openai2022moderation} filters plain toxicity, while Google's \textsc{Perspective API}~\cite{lees2022new} detects toxic content across languages. \textsc{BD-LLM}~\cite{distilling-detector} distills LLM knowledge to identify toxic prompts, and \textsc{WatchYourLanguage}~\cite{kumar2024watch} uses a reflection prompting mechanism with \textsc{GPT-4o} for detection.

\tool{} enhances detection by integrating both whitebox and blackbox approaches, utilizing LLM embeddings and an MLP classifier to provide a scalable and real-time solution for identifying toxic prompts.


\vspace{-0.65em}
\section{Conclusion}
\label{sec:conclusion}

\revision{In this work, we present \tool{}, a lightweight greybox method for efficiently detecting toxic prompts in LLMs. \tool{} leverages LLM-generated toxic concept prompts to create feature vectors and employs a classifier for prompt classification. Our evaluation on the latest LLMs, including LLama series and Gemma-2, demonstrates \tool{}'s high accuracy, low false positive rates, and superior performance compared to state-of-the-art methods. With a processing time of 0.078 seconds per prompt and the ability to train a detector in under five minutes, \tool{} is ideal for real-time applications. Future work will focus on adding interpretability and automated evaluation features to further enhance toxic prompt detection and ensure the safe use of LLMs.}
\vspace{-0.5em}
\section*{Acknowledgements}
We sincerely thank all the anonymous reviewers for their valuable feedback, which greatly contributed to the improvement of this paper. This research is jointly sponsored by the NSFC Program under Grants No. 62302304 and the ShanghaiTech Startup Funding. This research is supported by NTU College of Engineering CRP, Tier 3 Preparatory Grant 2023, and 10658 - MOE AcRF Tier 1: Call2/2023.

\clearpage
\bibliographystyle{acm}
\bibliography{ref}

\end{document}